\documentclass[]{aa}
\usepackage[normalem]{ulem}
\newcommand{\apropto}{\;
  \raise0.3ex\hbox{$\propto$\kern-0.75em\raise-1.1ex\hbox{$\sim$
  }}\;\hskip-2pt }

\newcommand{\lta}{\;
  \raise0.3ex\hbox{$<$\kern-0.75em\raise-1.1ex\hbox{$\sim$
  }}\;\hskip-2pt }
\newcommand{\gta}{\;
  \raise0.3ex\hbox{$>$\kern-0.75em\raise-1.1ex\hbox{$\sim$
  }}\;\hskip-2pt }

\usepackage[dvips]{graphicx}
\usepackage{graphicx}
\usepackage{ulem}
\usepackage{txfonts}
\usepackage{natbib}

\begin{document}
\title{The relation between magnetic and material arms in models for spiral galaxies }

\author{D. Moss\inst{1}\fnmsep
\thanks{Corresponding author, \email{moss@ma.man.ac.uk}}
R. Beck\inst{2}, D. Sokoloff\inst{3}, R. Stepanov\inst{4}, M.
Krause\inst{2} and T.\,G. Arshakian\inst{5}$^,$\inst{6}}
\titlerunning{Magnetic fields in spiral galaxies}
\authorrunning{D. Moss et al.}
\institute{School of Mathematics, University of Manchester,
Manchester M13 9PL, UK \and MPI f\"ur Radioastronomie, Auf dem
H\"ugel 69, 53121 Bonn, Germany \and Department of Physics, Moscow
State University, Russia \and Institute of Continuous Media
Mechanics, Korolyov str. 1, 614061 Perm, Russia, \and
I. Physikalisches Institut, Universit\"at zu K\"oln, Z\"ulpicher Str. 77, 50937
K\"oln, Germany \and Byurakan Astrophysical Observatory, Byurakan 378433, Armenia  and
Isaac Newton Institute of Chile, Armenian Branch}

\date{Received ?????; accepted ??????}

\abstract {Observations of polarized radio emission show that
large-scale (regular) magnetic fields in spiral galaxies are not
axisymmetric, but generally stronger in interarm regions. In some nearby
galaxies such as NGC~6946 they are organized in narrow magnetic arms
situated between the material spiral arms.} {The phenomenon of
magnetic arms and their relation to the optical spiral
arms (the material arms) call for an explanation in the framework
of galactic dynamo theory. Several possibilities have been suggested
but are not completely satisfactory; here we attempt a consistent
investigation.} {We use a 2D mean-field dynamo model in the no-$z$
approximation and add injections of small-scale magnetic field,
taken to result from supernova explosions, to represent the effects
of dynamo action on smaller scales.
This injection of small scale field
is situated along the spiral arms, where star-formation
mostly occurs.} {A straightforward explanation of magnetic arms as a
result of modulation of the dynamo mechanism by material arms
struggles to produce pronounced magnetic arms, at least with
realistic parameters, without introducing new effects such as a time
lag between Coriolis force and $\alpha$-effect. In contrast, by
taking into account explicitly the small-scale magnetic field  that
is injected into the arms by the action of the star forming regions that are concentrated there,
we can obtain dynamo  models with 
magnetic structures of various forms that can be compared with magnetic arms.  
These are rather
variable entities and their shape changes significantly on
timescales of a few $100$ Myr.
Properties of magnetic arms can be controlled
by changing the model parameters. In particular, a lower
injection rate of small-scale field makes the magnetic configuration
smoother and eliminates distinct magnetic arms.}
{We conclude that
magnetic arms  can be considered as coherent magnetic
structures generated by large-scale dynamo action, and
associated with spatially modulated small-scale magnetic fluctuations, caused by enhanced star formation rates within
the material arms.}


\titlerunning{Magnetic and material spiral arms in spiral galaxies}
\authorrunning{Moss et al.}
\maketitle

\section{Introduction}

Large-scale (regular) magnetic fields in the discs of spiral galaxies are
thought to be maintained by dynamo action, the joint action of
differential rotation and mirror asymmetric interstellar turbulence
associated with the entire galactic disc. According to
mean-field dynamo theory (e.g. Ruzmaikin et al. 1988; Beck et al.
1996), the dynamo organizes the field lines of the
large-scale magnetic field into spiral form. The tangent of the
pitch angle of the field is
given by the ratio of the radial and azimuthal
magnetic field components, and can also be expressed as a function of
the dynamo governing parameters.
The observed pitch
angles of the large-scale magnetic field are in the range between approximately
$10^\circ$ and $40^\circ$ (e.g. Fletcher 2010). Comparing these pitch angles
with the pitch angles of
the material (optical) spiral arms of such galaxies
shows that they are similar at least across significant parts of the
galactic discs; this cannot be understood
in the framework of the basic conventional form of the mean-field dynamo alone.

The field structure most readily generated by the mean-field dynamo
is axisymmetric (ASS, the $m=0$ mode). Radio observations of
nearby galaxies show, however, that while the total magnetic field
is stronger in the spiral arms, the
ordered magnetic field (traced by polarized emission)\footnote{Linearly polarized emission traces ``ordered'' magnetic
fields, which can be either ``regular'' (or large scale) magnetic fields
(preserving their direction over large scales) or ``anisotropic turbulent'' 
fields (with multiple field reversals within the telescope beam). 
To distinguish between these two components observationally, 
additional Faraday rotation data is needed.} is generally
stronger in the interarm regions (i.e. between the spiral arms).
This was first detected in the grand-design galaxy M~81 (Krause et al.
1989b) and has been confirmed in most nearby galaxies observed so far
at centimetre wavelengths(Beck 2005; Beck \& Wielebinski 2013).
A few galaxies with strong large-scale gas compression such as the density-wave galaxy
M~51 (Fletcher et al. 2011) and the barred galaxy NGC~1097 (Beck et al. 2005) show
additional polarized emission in the compression regions.

This result cannot be simply explained by stronger turbulence or enhanced Faraday depolarization within the
spiral arms but seems to indicate an
interaction between processes in the material spiral arms and the dynamo action
that generates large-scale field.

In particular, the magnetic configurations in some nearby spiral
galaxies are clearly organized into narrow  ``magnetic arms'', e.g. in IC~342 and
NGC~6946. Shifts between the magnetic and material arms are observed
in IC~342 (Krause 1993). There is a prominent spiral arm in the
south which splits into two branches.
The most spectacular example is provided by NGC~6946, a galaxy with
massive spiral arms, but without density-wave shock fronts. The
magnetic arms are clearly situated between material arms (Beck \&
Hoernes 1996, Beck 2007). The magnetic arms in NGC~6946 are most prominent
in polarized emission. Strong Faraday rotation in the magnetic arms shows
that the field is large-scale (regular). On the other hand, the random
magnetic field (traced by the unpolarized emission) is strongest
in the material arms.

There have been several attempts to introduce the
interaction between spiral arms and magnetic arms.
A direct effect of a density wave is a
modulation of the axisymmetric velocity distribution of the gas in a
galactic disk. The effect of this modulation on the galactic dynamo
was examined by Chiba \& Tosa (1990) (see also Moss 1998) who found conditions for
parametric resonance to occur, the so-called swing excitation of galactic
magnetic fields.
However, the rather special conditions that have to be satisfied for such
resonances to occur suggest that they are unlikely to be of major
physical importance.

Shukurov (1998) suggested that -- as a result of the gas density contrast --
random gas velocities in the material arms are larger than those in
the interarm space, and that they destroy large-scale magnetic
fields within the arms. Magnetic fields may survive in the interarm
space, in the form of magnetic arms situated between the material
arms, provided that dynamo action is not too strong.

The main problem here is that our knowledge concerning the
arm/interarm contrast in dynamo governing parameters is very limited
(Shukurov \& Sokoloff 1998; Moss 1998; Rohde et al. 1999; Chamandy
et al. 2013a). As we only have snapshots of the structure of few
nearby galaxies it is not clear how general the example of NGC~6946
is. It might also be connected with the material arms rotating
 with an angular
velocity that is different from that of the gas in the disc, so that
sometimes both types of arms coincide and sometimes they are
displaced.

It looks a priori plausible that by modifying the
non-axisymmetric distribution of the dynamo-governing parameters, a
magnetic arm configuration of the desired type might be obtained. In
fact,
Moss (1998) and Elstner et al.
(2000) considered the enhancement of the turbulent diffusivity and
the $\alpha$-effect in the spiral arms due to density waves, and
effects on the evolution of magnetic arms.
In addition, Kulpa-Dybel et al. (2011) claim that some interlacing
of magnetic and material arms can occur from the difference in
angular velocity between the arm pattern and the gas.

However, the available observational information and theoretical
ideas presented up to now are insufficient to isolate a completely
plausible explanation
for the phenomenon of magnetic arms.
For example, while the parametric resonance mechanism can produce
something like magnetic arms, rather special combinations of
galactic parameters are necessary to produce significant effects.
(We note that the problems
associated with material arms themselves are not yet completely
clear).
Chamandy et al. (2013a) do produce magnetic arms by a combination of
a spatial modulation of the alpha effect and temporal nonlocality -- see
also Chamandy et al. (2013b).

In order to limit the variety of the models, we consider a
dynamo model based on a particular parametrization of the
arm--interarm contrast in the dynamo-governing parameters, suggested
by Shukurov \& Sokoloff (1998).
(We acknowledge that any such estimates are subject to substantial
uncertainty.)
Then we focus our attention on a further idea
that mean-field dynamos and large-scale magnetic fields exist in the
presence of small-scale magnetic fields and various local
distortions of a smooth dynamo action. We demonstrate that such
small-scale mechanisms, when modulated by a spiral structure, are
able to produce magnetic arms. The model used here is a modification
of that of Moss et al. (2012).

Before moving to specific modelling we briefly present here the
leading idea of this paper.
We first confirm that it is possible to obtain magnetic arms
situated between the material arms, just by
spatial variation of the dynamo governing parameters (Sect.~3).
This mechanism is in some ways not completely satisfactory,  and we deduce
that it is desirable to find another mechanism leading to magnetic arm
formation. We suggest such a  mechanism, based on the idea that
the spatial scale of magnetic fluctuations
associated with mean-field galactic dynamo is in fact not very much smaller
than the spatial scale of the large-scale magnetic
field, and that the contribution of the fluctuations 
to the evolution of the mean-field should be
taken into account. The importance of fluctuations for the evolution
of the mean-field field
was stressed in general form by Hoyng (1988), and we apply the general idea
in the specific framework of galactic dynamos.
Following this approach we include
these fluctuations in the conventional mean-field equations.
We appreciate that such extension
of the mean-field equations deserves further verification at
a fundamental level; however we do not consider this as a goal of
the current paper.

We exploit a plausible assumption that
small-scale magnetic field injection is enhanced in the material arms,
because of the higher star formation rate and increased occurrence
of star forming regions there.
In order to illuminate
the effect suggested we deliberately assume that all other dynamo governing
parameters
are independent of azimuth. Of course, we appreciate that in practice variations in the injection rate
will very likely be associated with some variations of the other dynamo governing parameters,
and do not argue against the idea that such mechanisms can also contribute
to the formation of magnetic structures.
We emphasize that our aim in this paper is not to produce a comprehensive
modelling of magnetic fields in spiral galaxies, including all possible relevant
effects. Rather we attempt to isolate the effects of a novel mechanism, which
we believe can contribute to the formation of regular interarm fields in some
instances.

\section{The dynamo model}

We use here a plausible simplification of the 2D mean-field galactic
dynamo in the form of the ``no-$z$'' model (e.g. Subramanian \&
Mestel 1993; Moss 1995), which restricts modelling to quantities
which are accessible observationally and make the numerical
implementation easily affordable.
For the sake of clarity we briefly reproduce the relevant equations from
\cite{metal12}.

The code solves in the $\alpha\omega$ approximation explicitly for
the field components parallel to the disc plane, while the component
perpendicular to this plane (i.e. in the $z$-direction) is given by
the solenoidality condition. The even (quadrupole-like) magnetic
field parity with respect to the disc plane is assumed. The field
components parallel to the plane are considered as mid-plane values,
or as a form of vertical average through the disc. The key
parameters are the aspect ratio $\lambda=h_0/R$, where $h_0$
corresponds to the semi-thickness of the warm gas disc and $R$ is
its radius, and the  dynamo numbers $R_\alpha=\alpha_0 h_0/\eta,
R_\omega=\Omega_0 h_0^2/\eta$. We allow the disc semi-thickness
$h=h(r)$ to vary with radius (see below), and $h_0=h(0)$ is our
reference value. $\lambda$ must be a small parameter. $\eta$ is the
turbulent diffusivity, assumed uniform, and $\alpha_0, \Omega_0$ are
typical values of  the $\alpha$-coefficient and angular velocity
respectively. Thus the dynamo equations become  in cylindrical polar
coordinates $(r, \phi, z)$
\begin{eqnarray}
\frac{\partial B_r}{\partial t} = -R_\alpha\frac{h(r)}{h_0} B_\phi-\frac{\pi^2}{4}\left(\frac{h(r)}{h_0}\right)^2
B_r +\\ \nonumber
+\lambda^2\left(\frac{\partial}{\partial
r}\left[\frac{1}{r}\frac{\partial}{\partial
r}(rB_r)\right]+
\frac{1}{r^2}\frac{\partial^2B_r}{\partial\phi^2}-\frac{2}{r^2}\frac{\partial
B_\phi}{\partial\phi}\right) , \label{evolBr}
\end{eqnarray}
\begin{eqnarray}
\frac{\partial B_\phi}{\partial t} = R_\omega r
B_r\frac{d\Omega}{dr}-R_\omega\Omega\frac{\partial B_\phi}{\partial
\phi}-\frac{\pi^2}{4} B_\phi\left(\frac{h(r)}{h_0}\right)^2 +\\ \nonumber
+\lambda^2\left(\frac{\partial}{\partial
r}\left[\frac{1}{r}\frac{\partial}{\partial r}(rB_\phi)\right]
+\frac{1}{r^2}\frac{B_\phi^2}{\partial
\phi^2}-\frac{2}{r^2}\frac{\partial B_r}{\partial \phi}\right),
\label{evolBphi}
\end{eqnarray}
where $z$ does not appear explicitly. Here a flared disc  with
semi-thickness
\begin{equation}
h=h_0 (1+{r}/{r_h})^{1/2}
\end{equation}
is assumed, with $h_0, r_h$ being constants. The factors $h(r)/h_0$
are introduced in Eqs.~(\ref{evolBr}), (\ref{evolBphi})
to allow for the variation in disc height with
radius, following Ruzmaikin et al. (1988).  (We note that Lazio \& Cordes (1998) argue that the Milky Way disc
is flat; there is no reliable evidence concerning disc
flaring in external galaxies.  Limited
experimentation suggests that our results do not depend significantly
on this assumption, within the uncertainty in the dynamo parameters.)
This equation has been calibrated by introduction of the factors
$\pi^2/4$ in the vertical diffusion terms. In principle, in the
$\alpha\omega$  approximation the parameters $R_\alpha, R_\omega$
can be combined into a single dynamo number $D=R_\alpha R_\omega$,
but we choose to keep them separate. Length, time and magnetic field
are non-dimensionalized in units of $R$, $h_0^2/\eta$ and the
equipartition field strength $B_{\rm eq}$, respectively.
$\alpha_0=\alpha_0(r)$ is taken proportional to $h(r)\Omega(r)$,
and is unaffected by the arms. This, together with the presence of uniform
turbulent diffusivity, implies the presence of a background level
of turbulence throughout the disc.
A naive
algebraic $\alpha$-quenching nonlinearity is assumed,
$\alpha=\alpha_0/(1+B^2/B_{\rm eq}^2)$, where $B_{\rm eq}$ is the
strength of the equipartition field in the general disc environment.

We take the same rotation law as used in Moss et al. (2012), namely

$$r\frac{d\Omega}{dr}= \quad \quad \quad \quad \quad \quad\quad \quad \quad \quad \quad \quad \quad \quad \quad \quad \quad \quad\quad $$
\begin{equation}
=\Omega_0\left(-\frac{1}{r R_{\rm gal}}\tanh(\frac{r R_{\rm gal}}{r_0})+\frac{1}{r_0\cosh^2(r R_{\rm gal}/r_0)}\right),
\nonumber
\label{rot}
\end{equation}
where $r_0$ corresponds to the turnover radius for the rotational
velocity: we take $r_0=0.2$.

\begin{figure*}
(a)\includegraphics[height=0.3\textwidth]{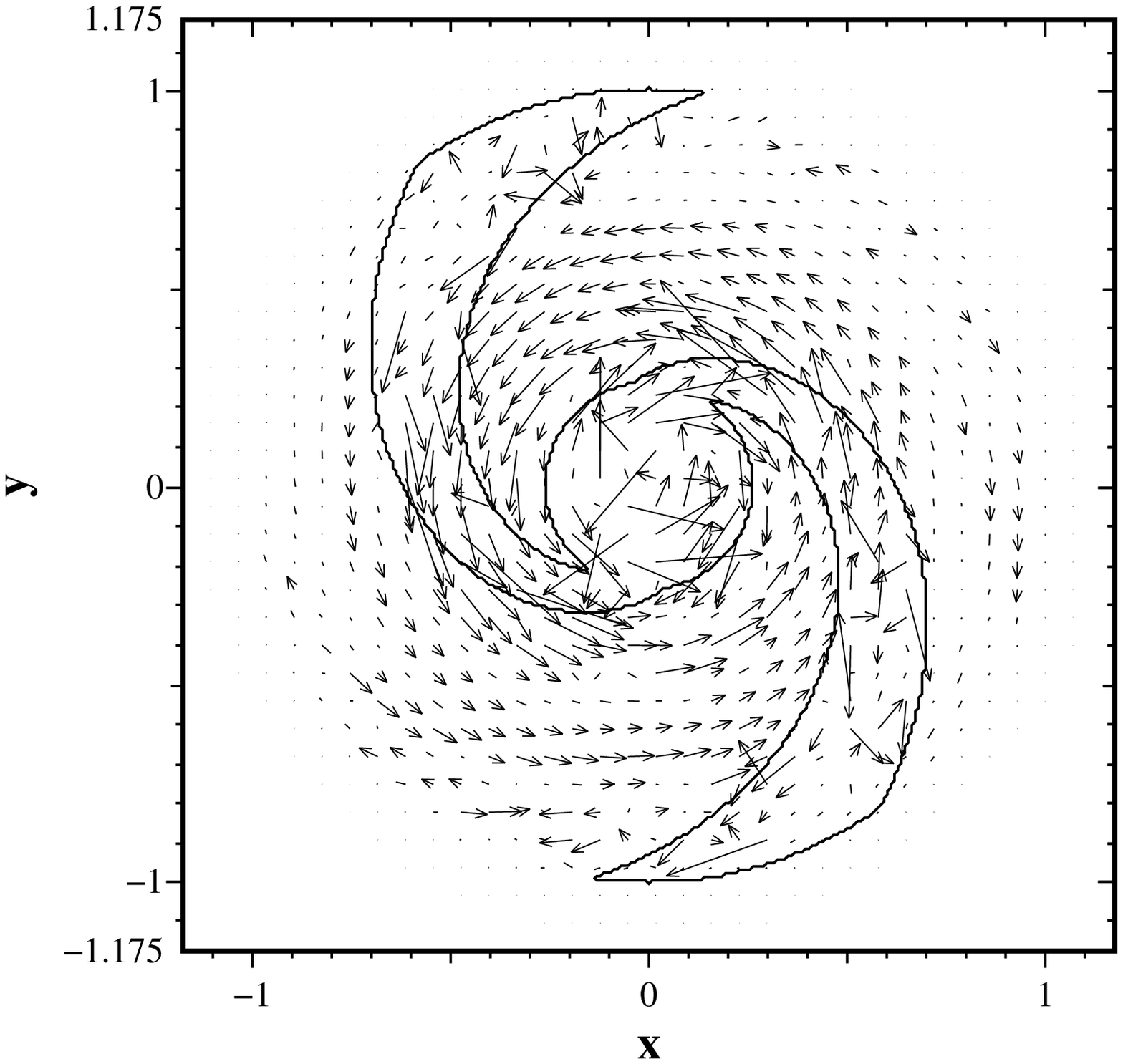}
(b)\includegraphics[height=0.3\textwidth]{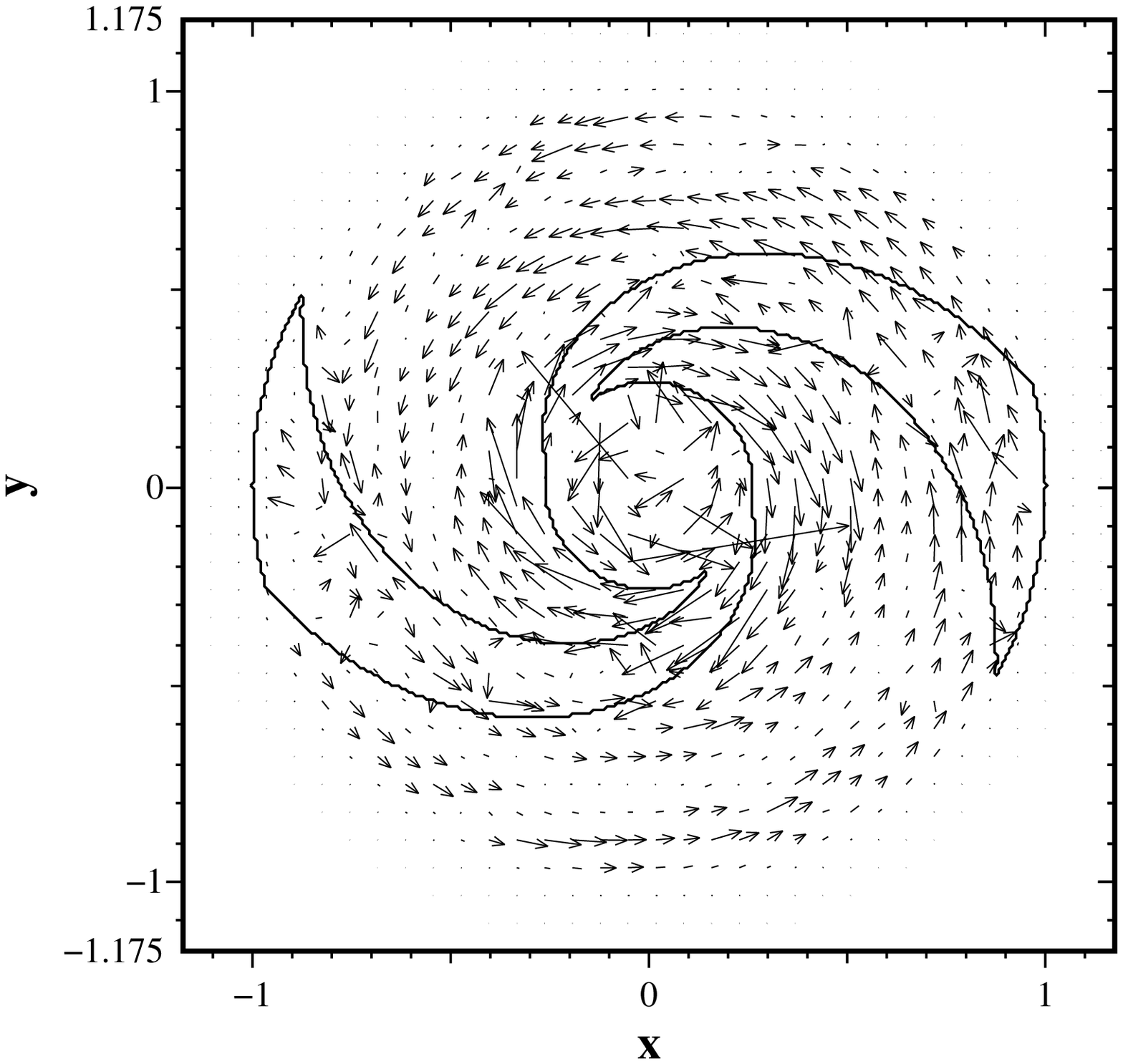}
(c)\includegraphics[height=0.3\textwidth]{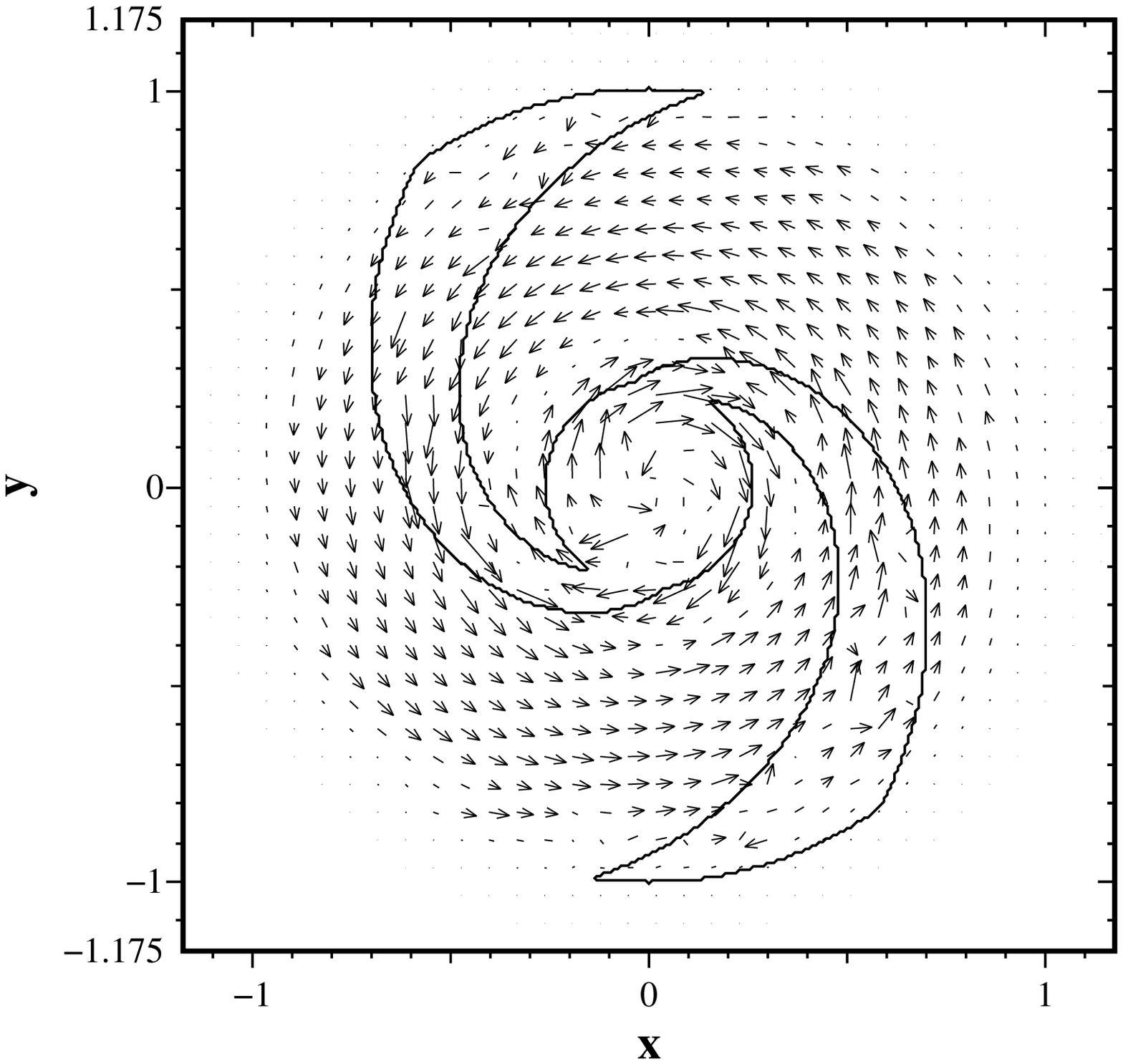}\\
(d)\includegraphics[height=0.3\textwidth]{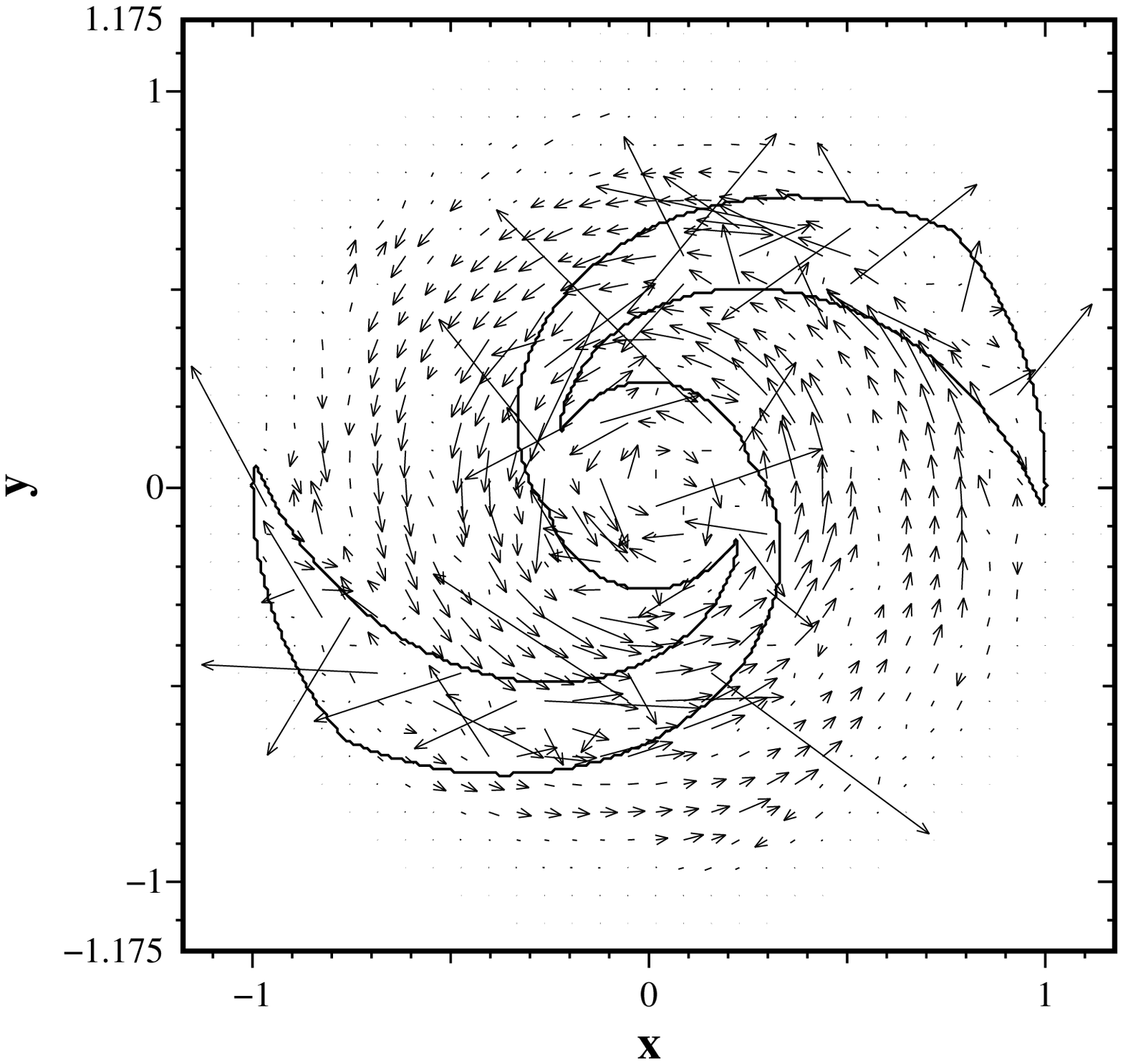}
(e)\includegraphics[height=0.3\textwidth]{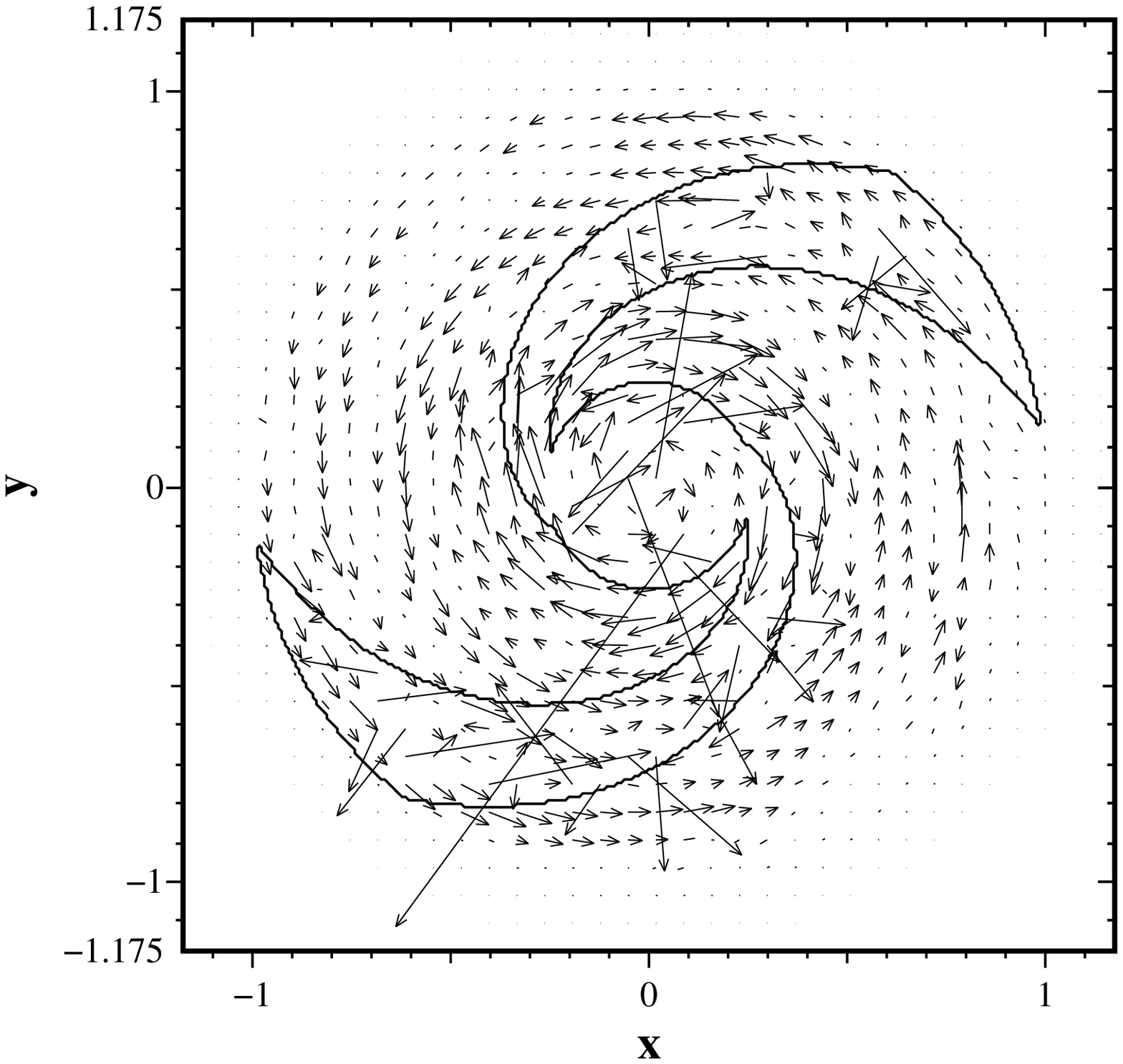}
(f)\includegraphics[height=0.3\textwidth]{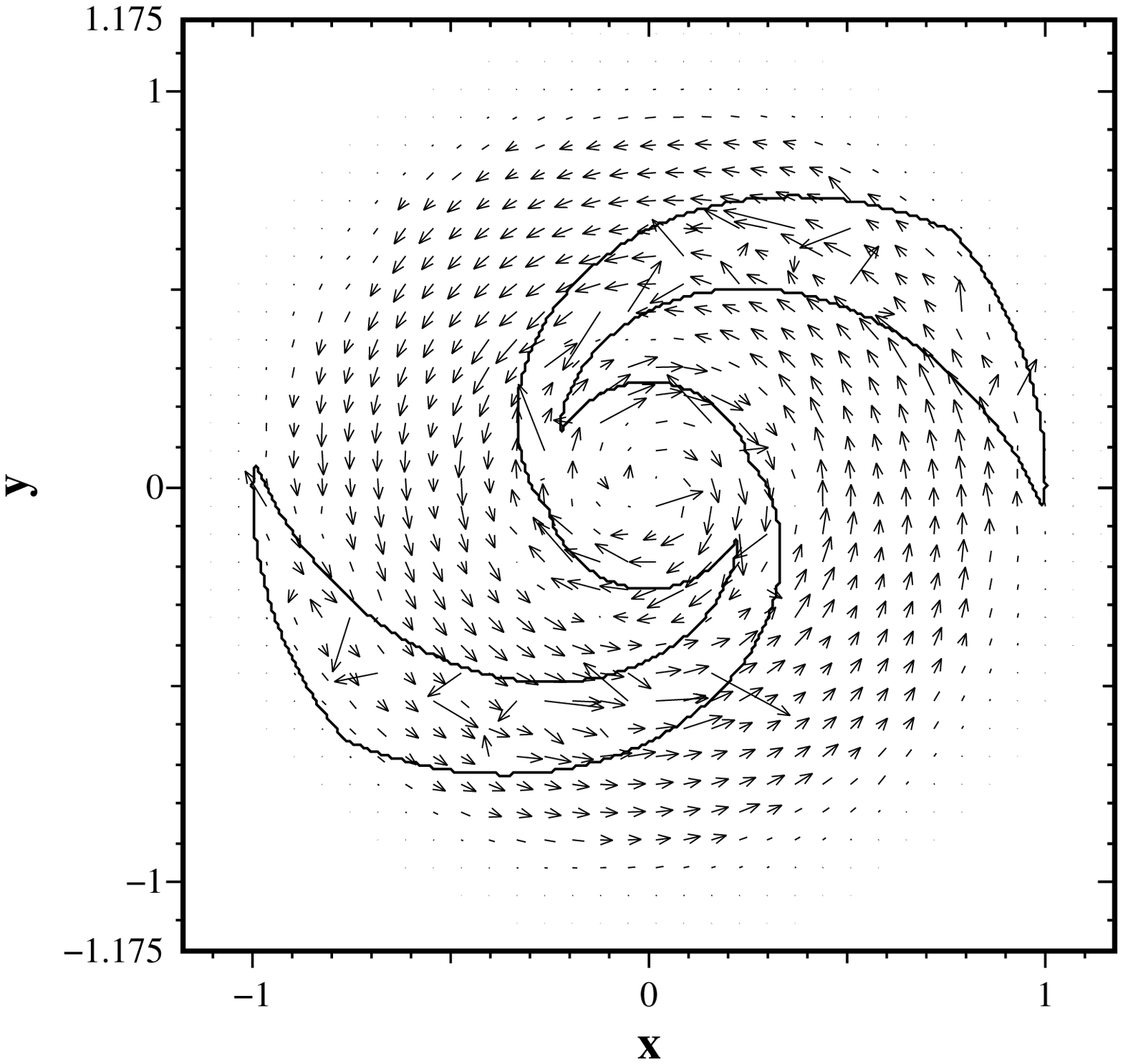}\\
(g)\includegraphics[height=0.3\textwidth]{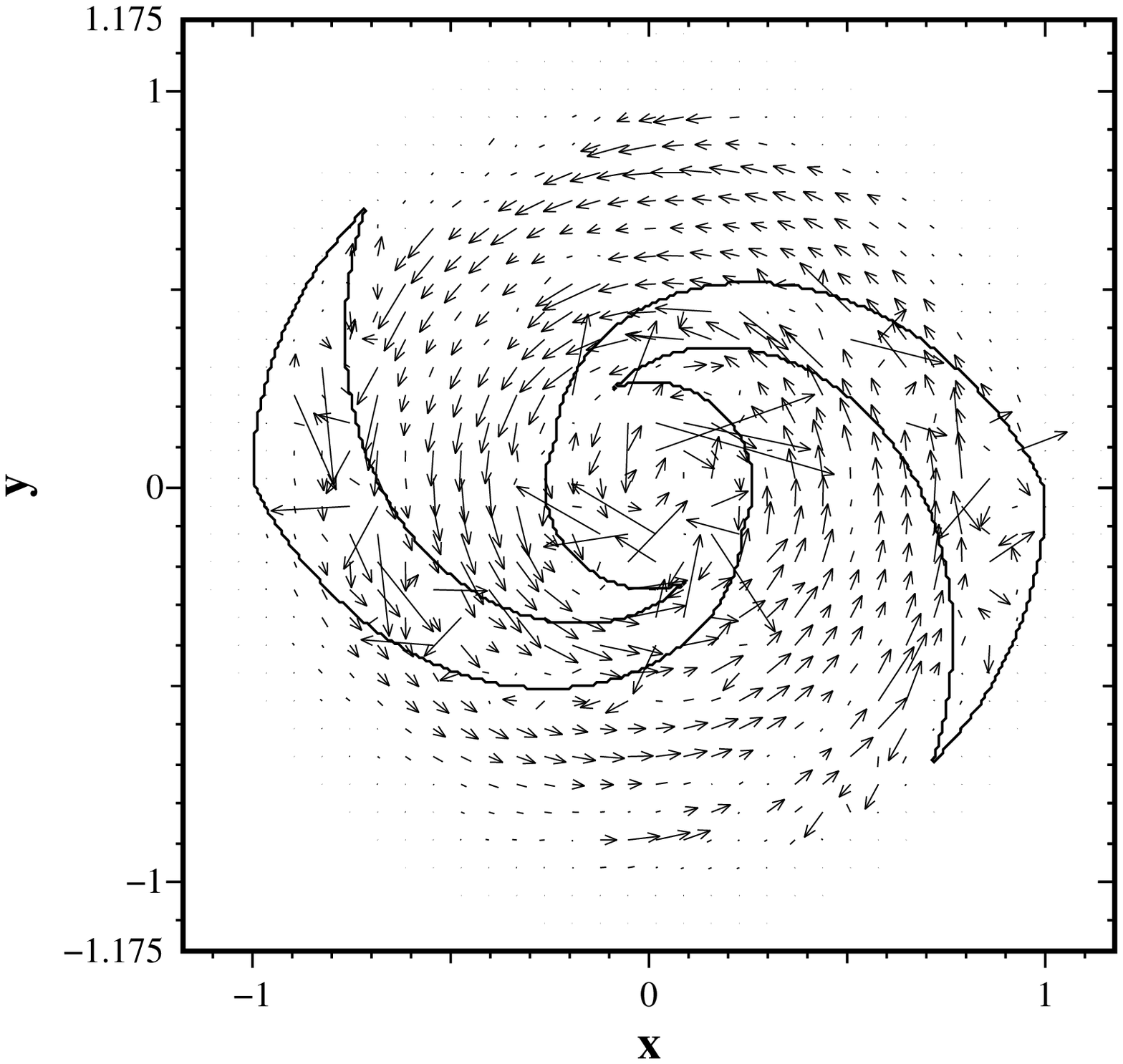}
(h)\includegraphics[height=0.3\textwidth]{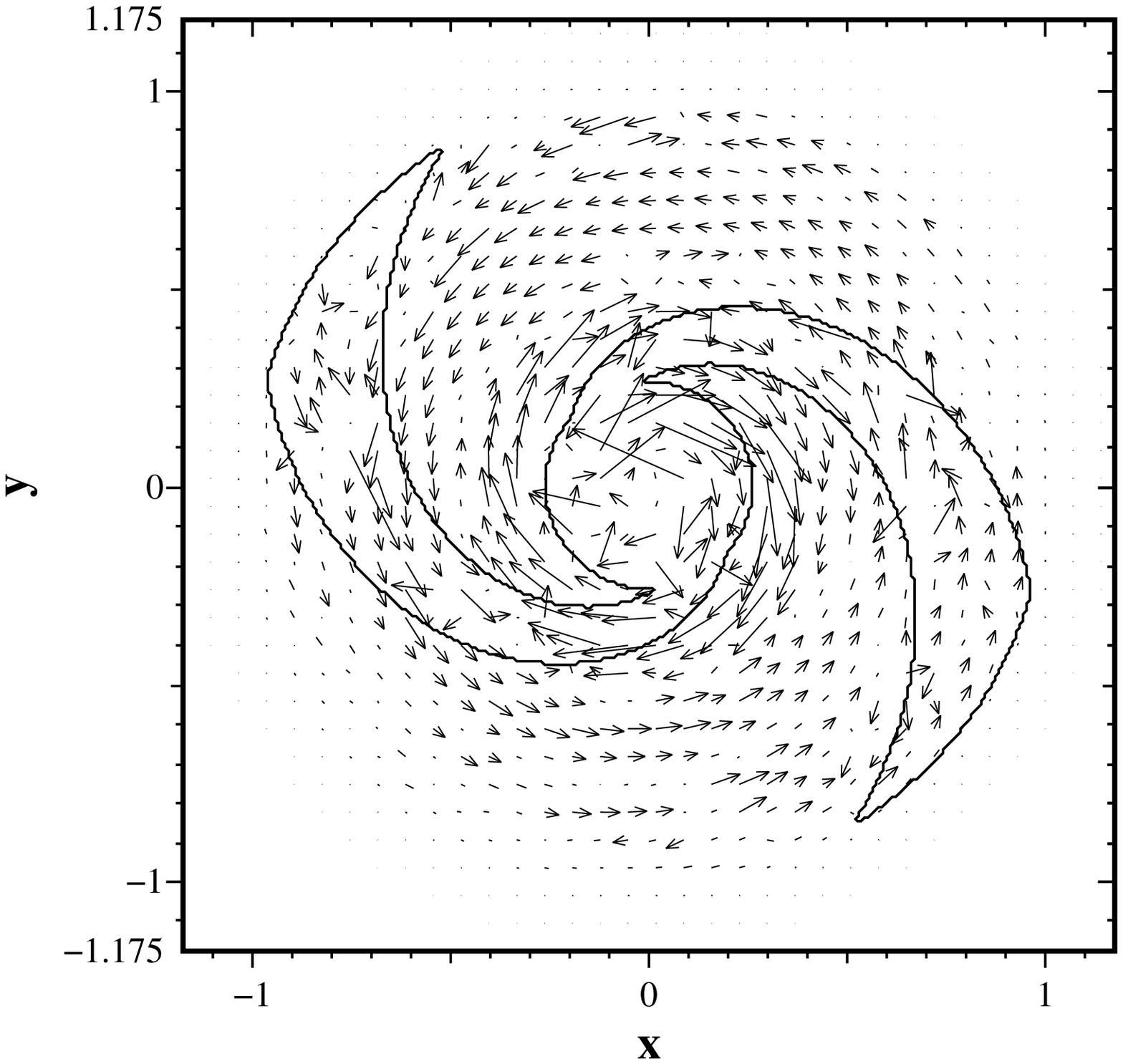}
(i)\includegraphics[height=0.3\textwidth]{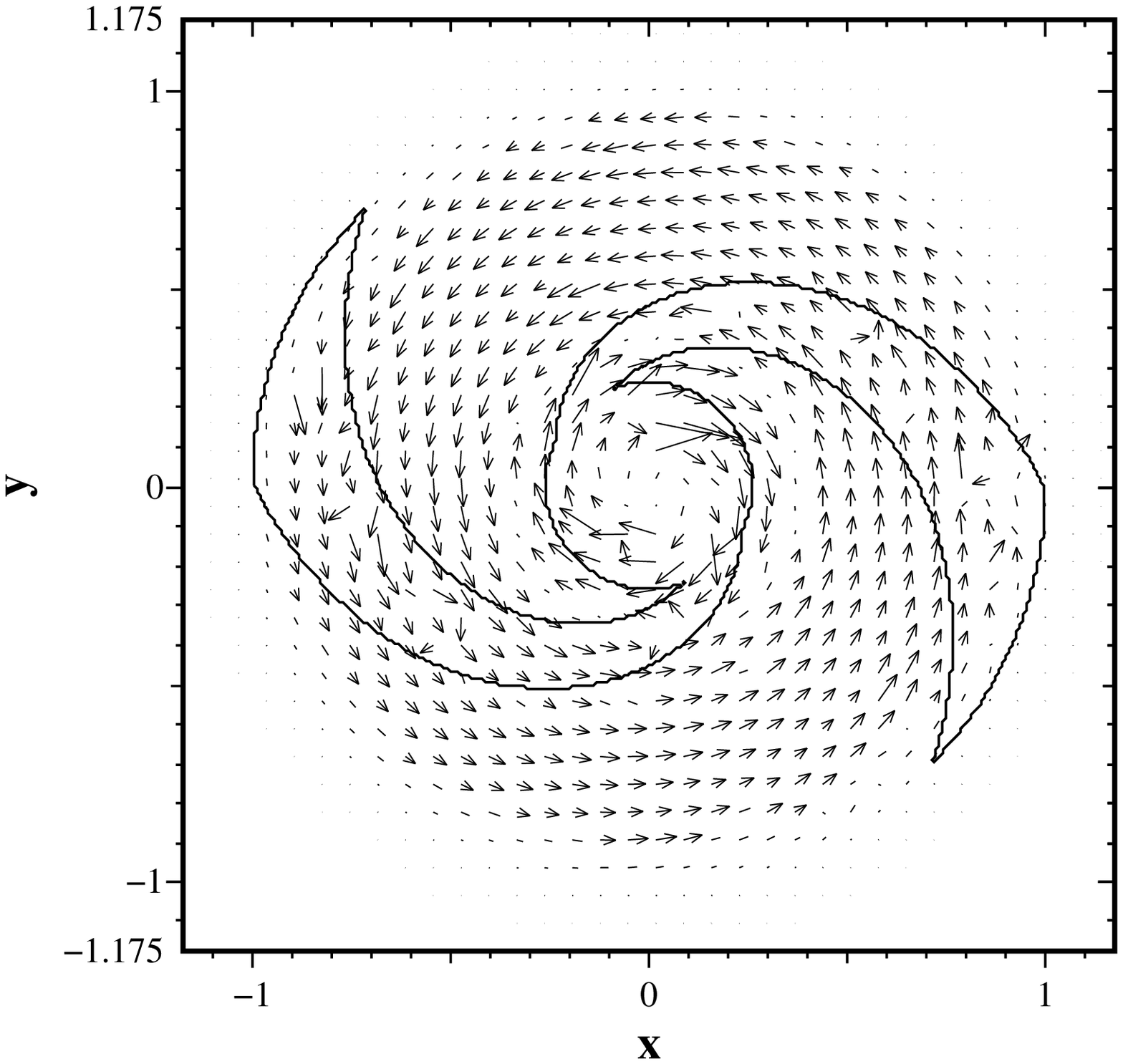}\\
(j)\includegraphics[height=0.3\textwidth]{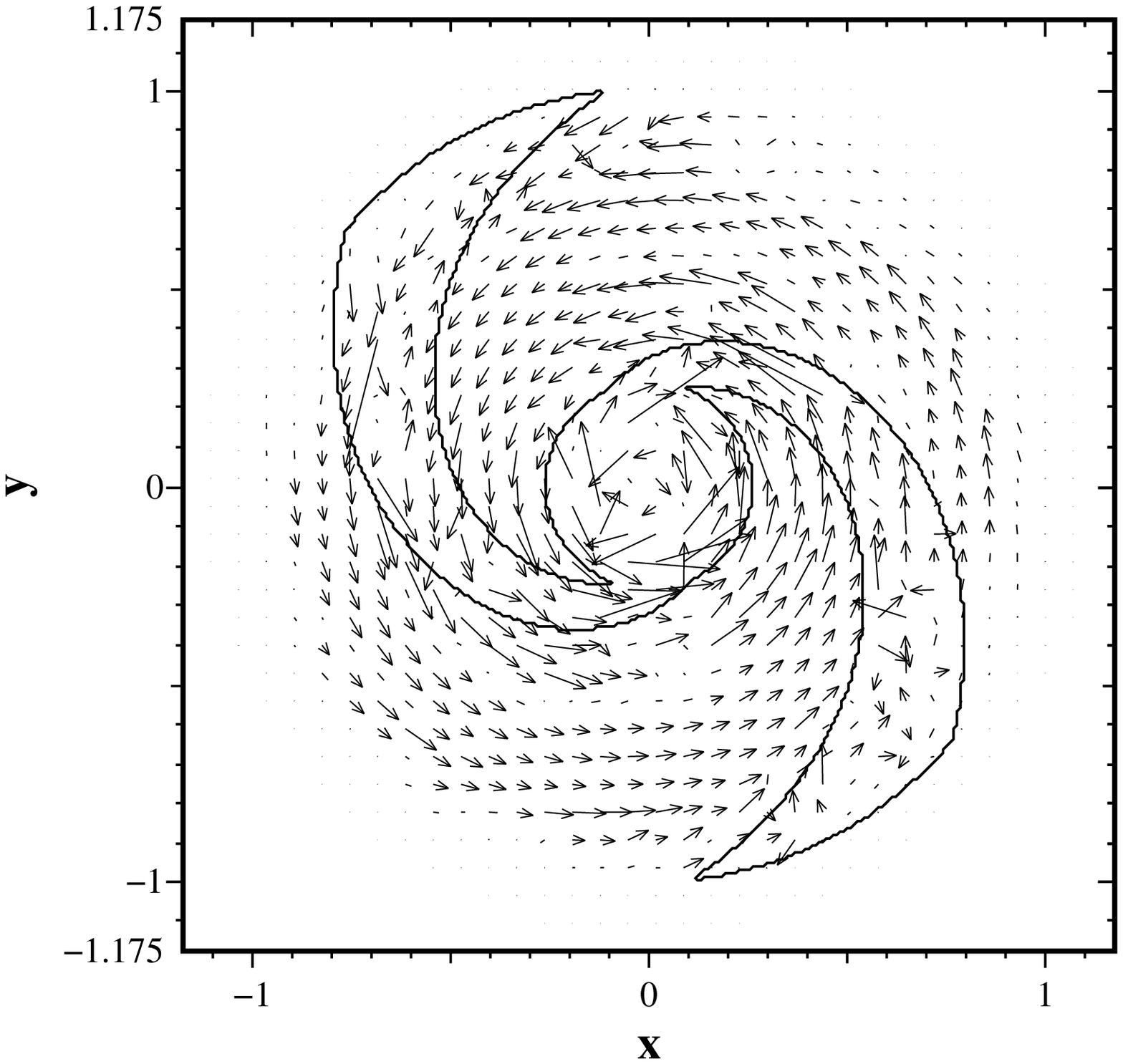}
(k)\includegraphics[height=0.3\textwidth]{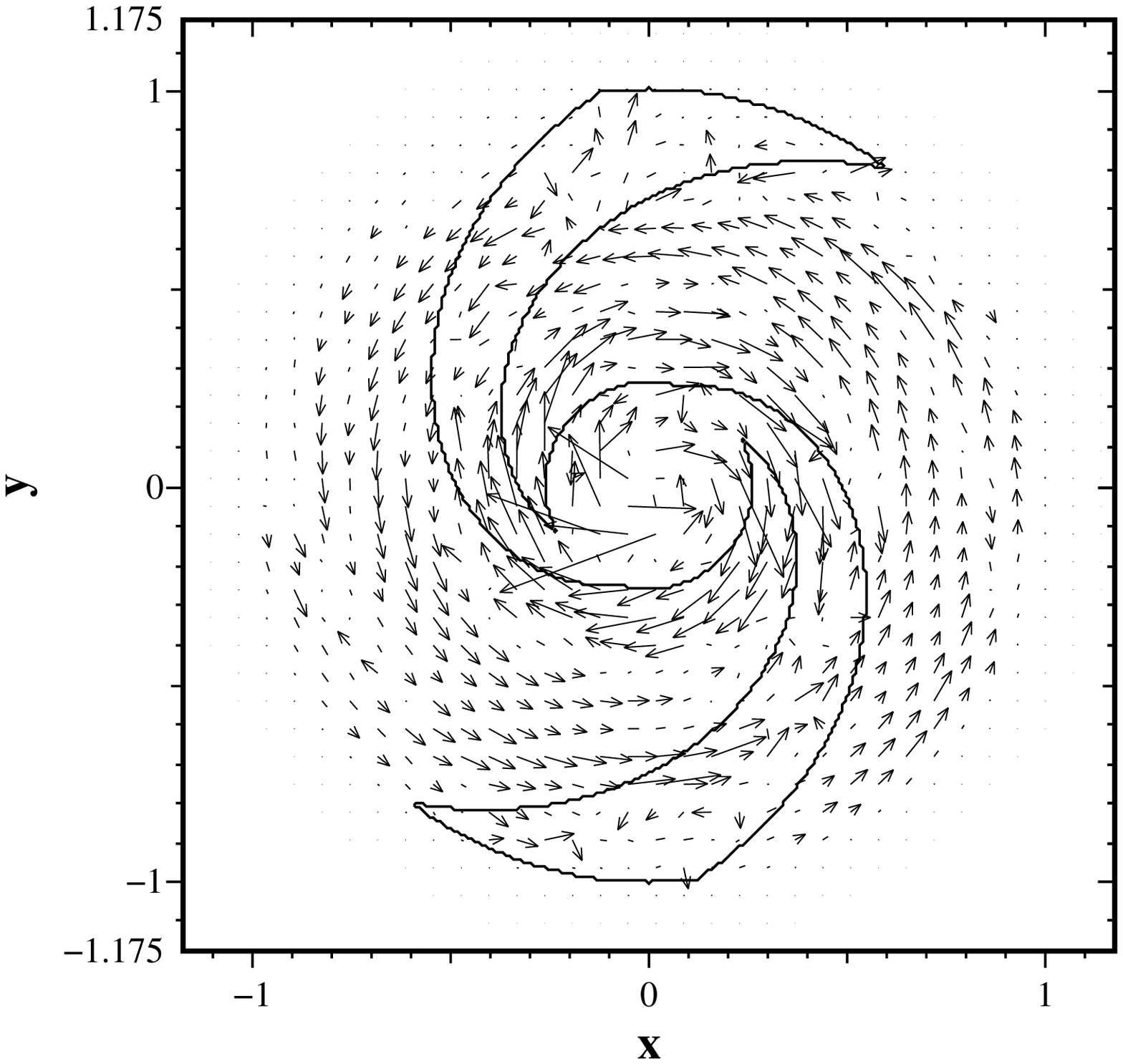}
(l)\includegraphics[height=0.3\textwidth]{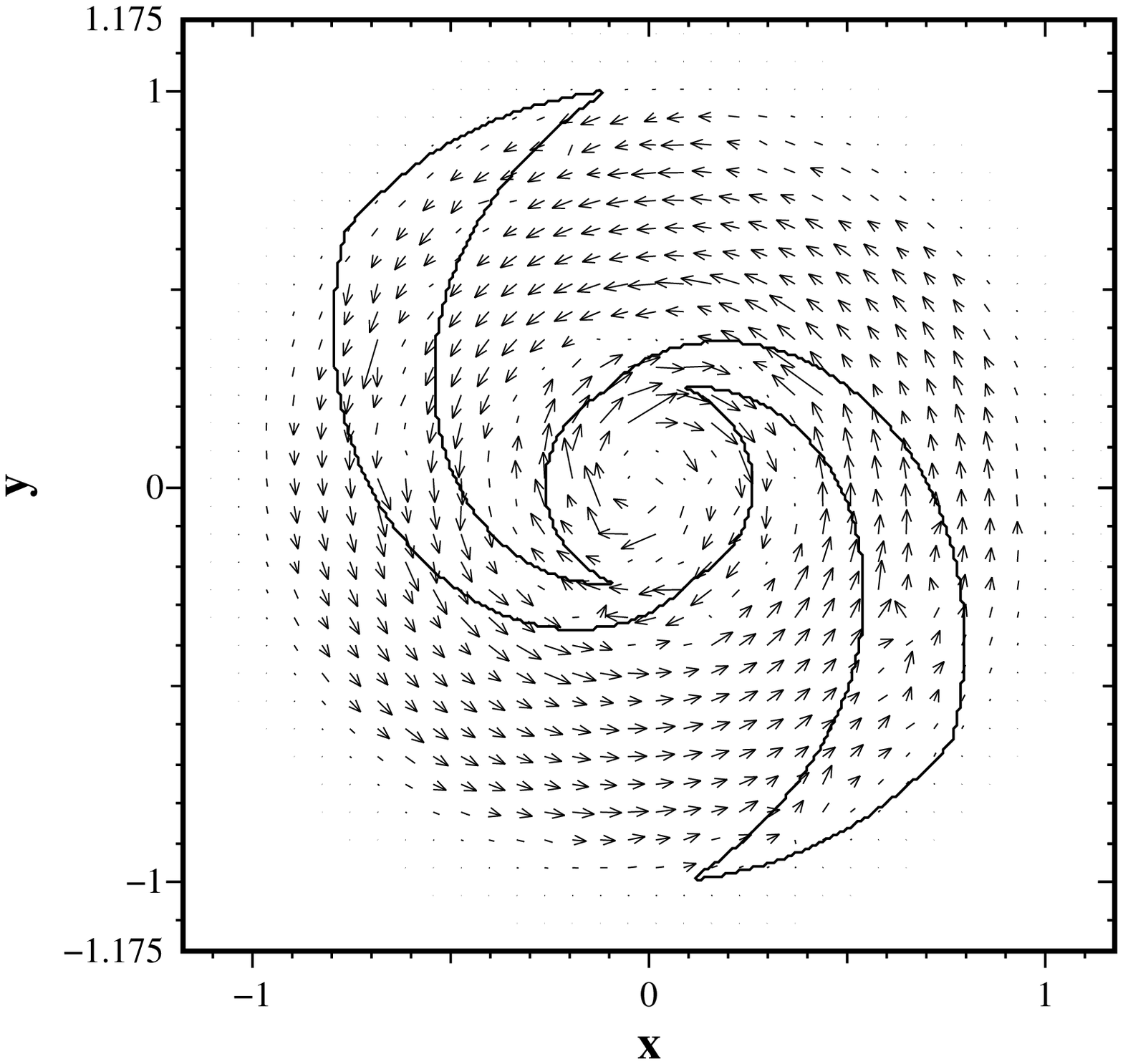}
\caption{Evolution of magnetic field configurations. Left-hand
column is model 75, middle column model 76, right-hand column model
77. Models (a), (b), (c) (first row) at approximate time $11.7$~Gyr;
models (d), (e), (f) (second row) at time $12.5$~Gyr; models (g),
(h), (i) (third row) at $12.9$~Gyr; models (j), (k), (l) (bottom
row) at approximately $13.2$~Gyr (i.e.  "now"). For model 76 the rotation parameter $R_\omega$
has been increased from that of model 75, and in model 77 the magnitude of the injected field,
$B_{\rm inj0}$, has been decreased from that used in model 75. The continuous contours delineate the regions
("arms") where field is injected;  these rotate rigidly,
with pattern speed such that the corotation radius is at
fractional radius $r\approx 0.7$.} \label{evol}
\end{figure*}

\begin{figure*}
(a)\includegraphics[height=0.3\textwidth]{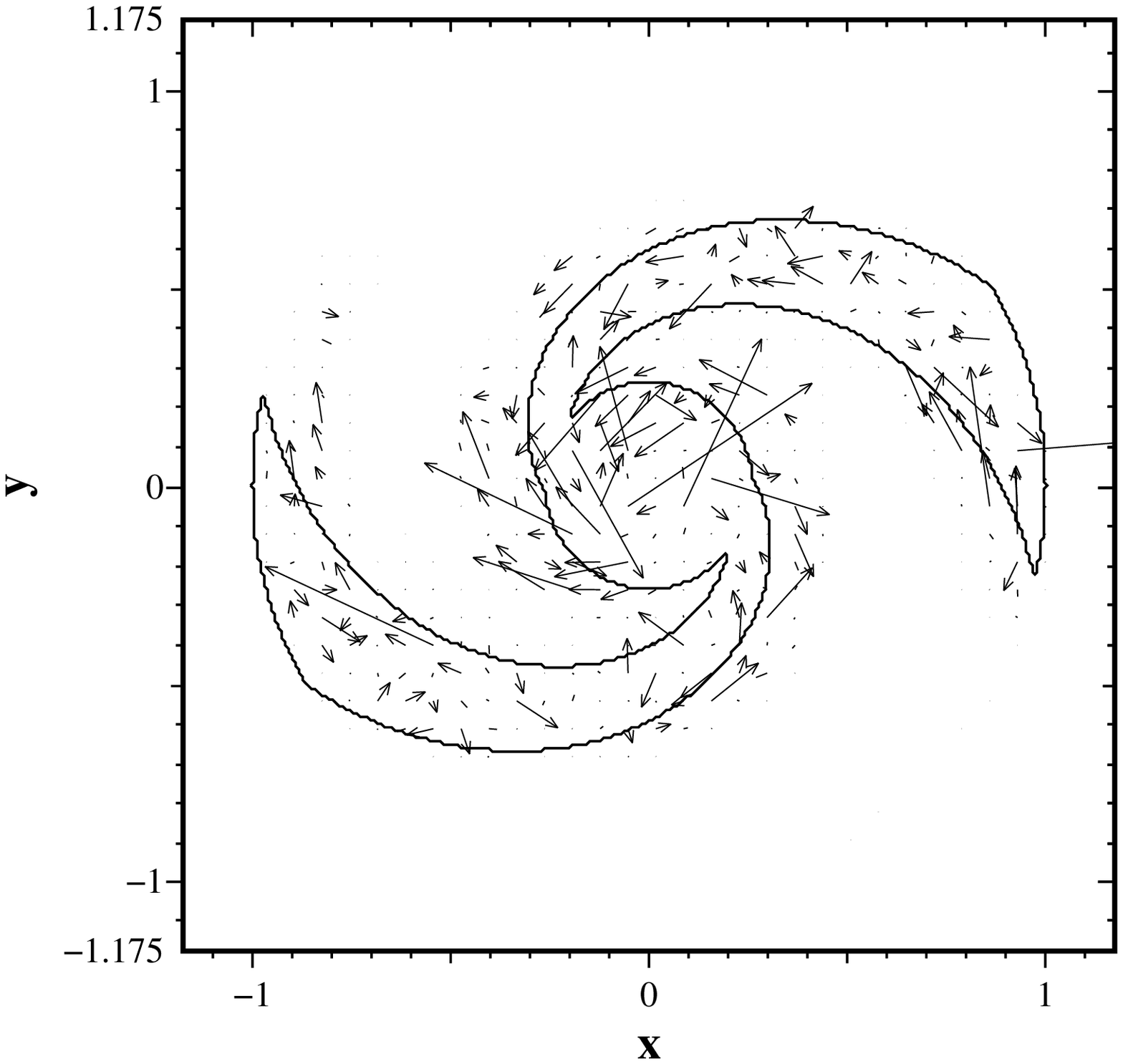}
(b)\includegraphics[height=0.3\textwidth]{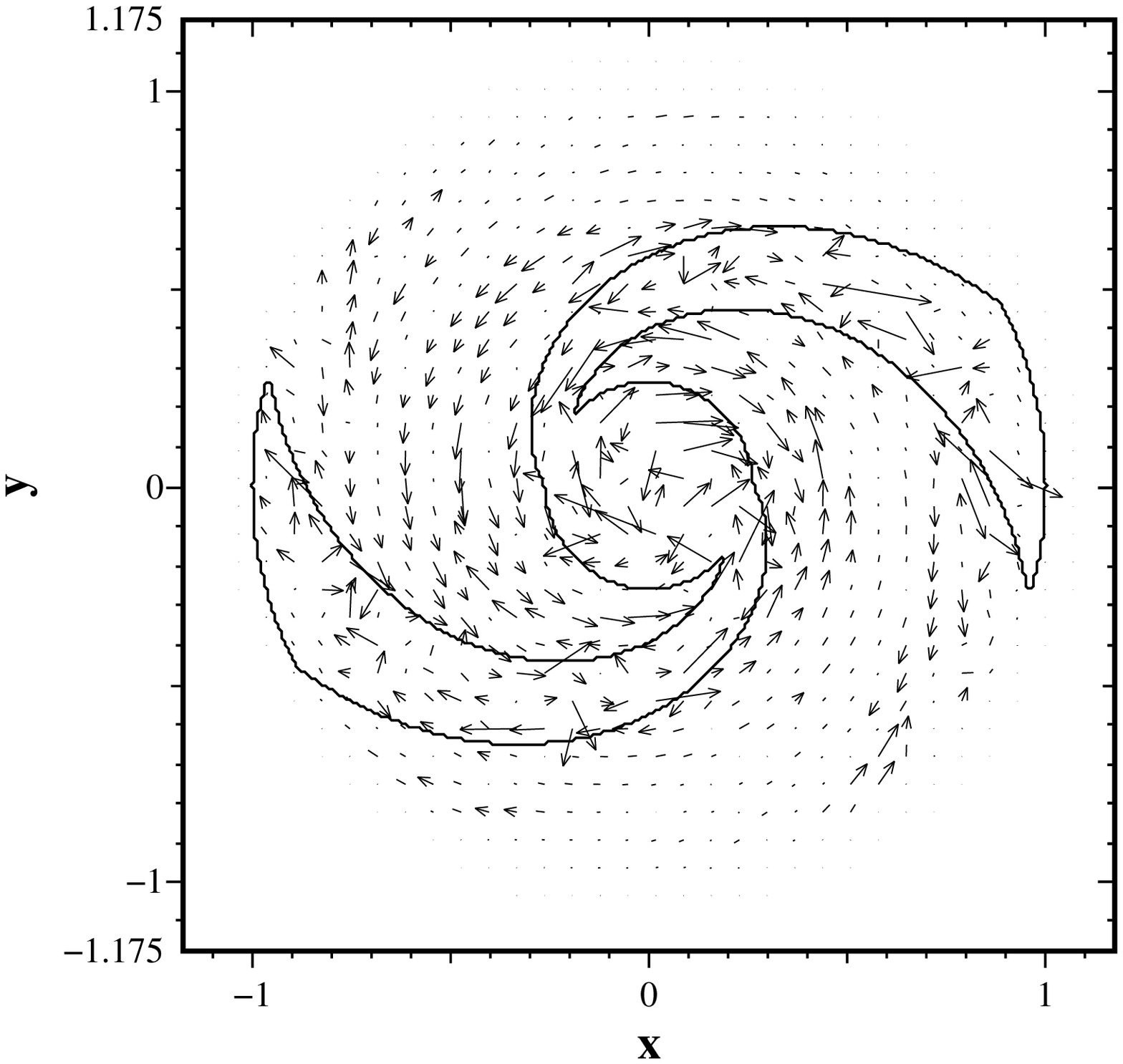}
(c)\includegraphics[height=0.3\textwidth]{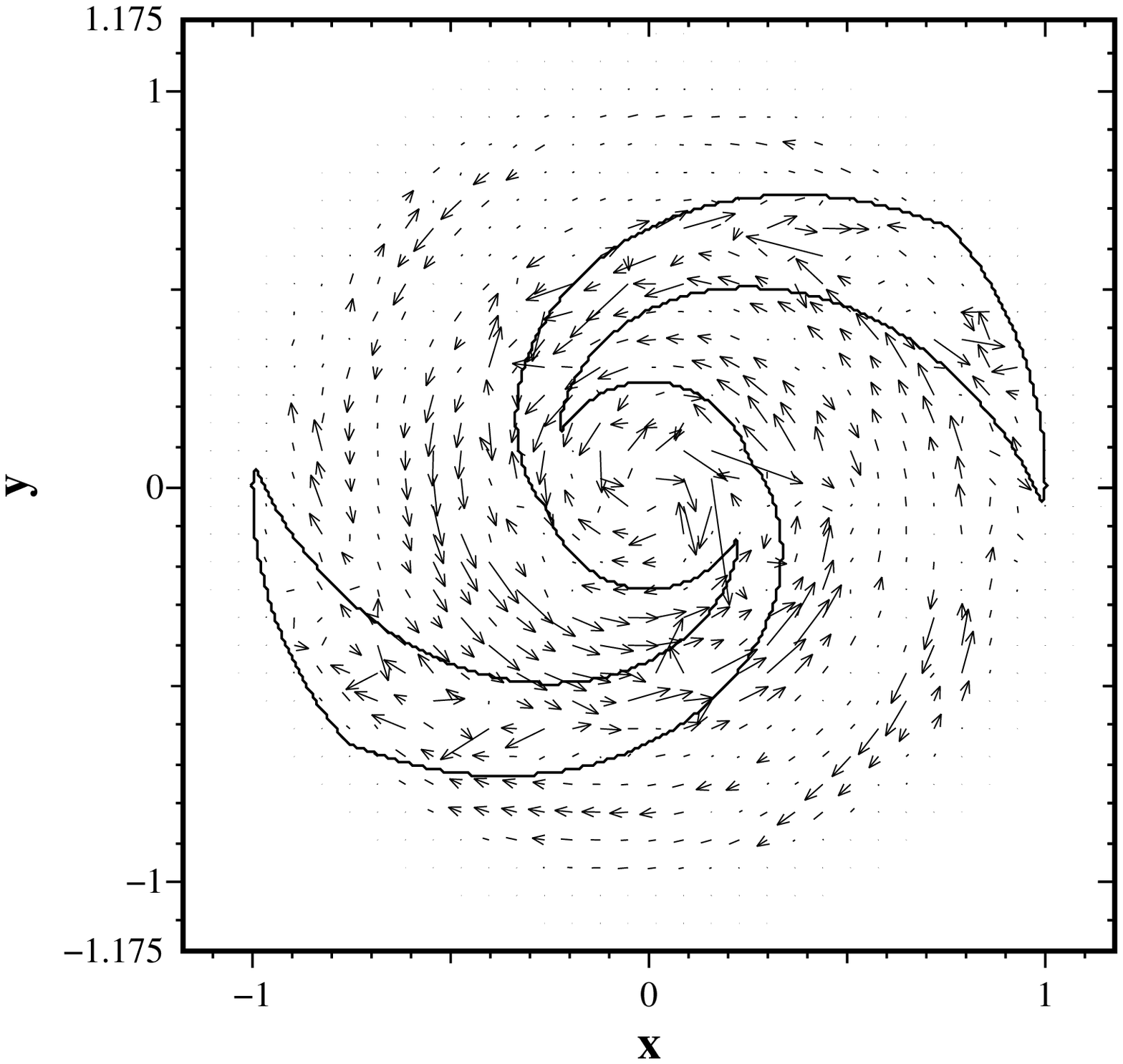}\\
(d)\includegraphics[height=0.3\textwidth]{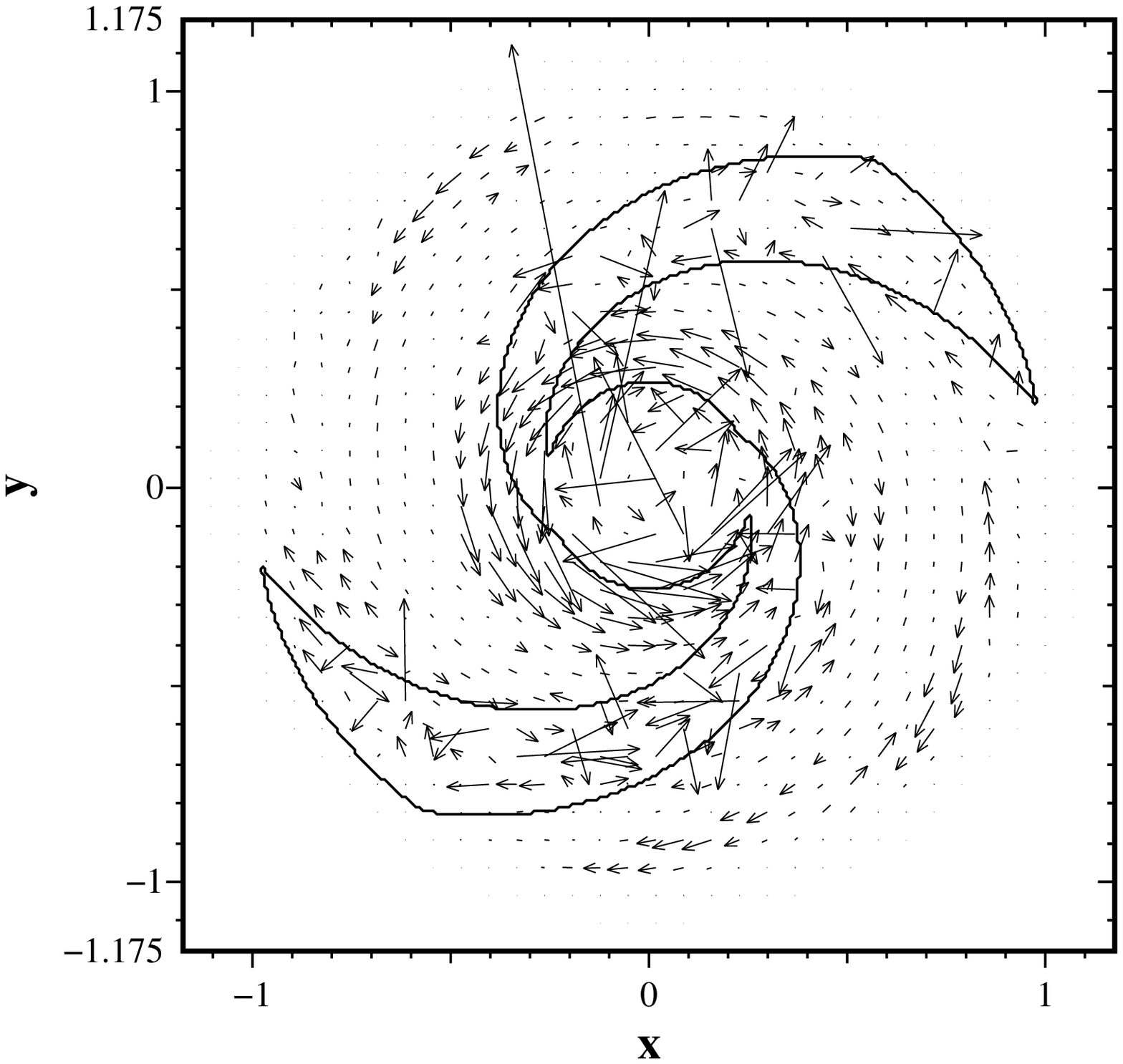}
(e)\includegraphics[height=0.3\textwidth]{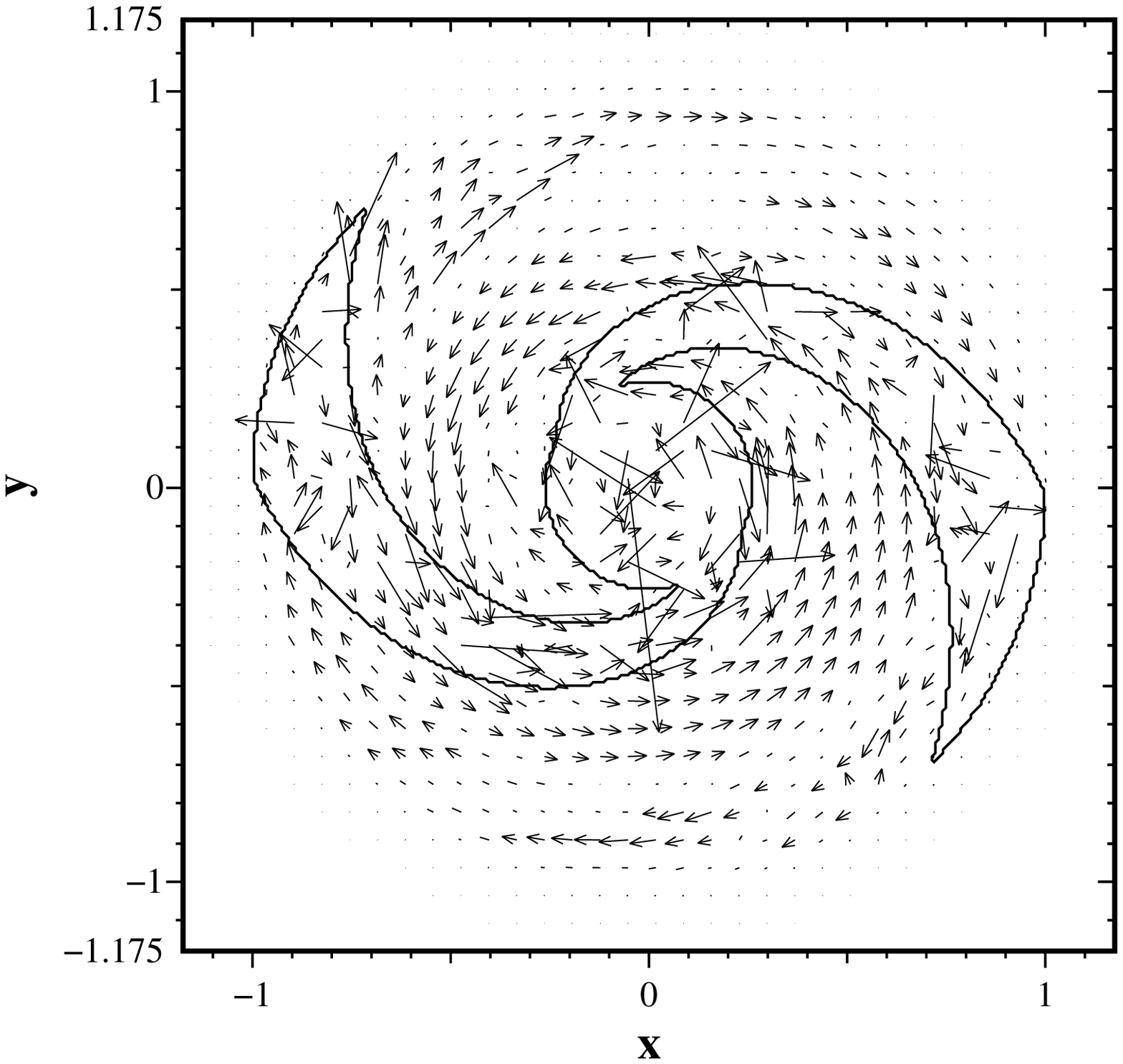}\\
\caption{The early evolution of the magnetic field in model 75.
 Models (a), (b), (c) (first row) at approximate time $0.078, 0.78, 2.34$~Gyr
respectively;
models (d), (e) (second row) at times $3.9, 7.8$~Gyr.
The continuous contours delineate the regions
("arms") where field is injected;  these rotate rigidly,
with pattern speed such that the corotation radius is at
fractional radius $r\approx 0.7$. As time proceeds, the large-scale organization of the field increases -- see also the first column of \protect Fig.~\ref{evol}.} \label{early}
\end{figure*}

The alpha effect is assumed to depend on angular velocity and disc
thickness, $\alpha\propto \Omega h^{-1}$, whereas $B_{\rm eq}$ is
assumed to be uniform because our restricted knowledge
of physical conditions
gives no secure basis for more
realistic assumptions. We continue our simulations to a time
corresponding to the present day where knowledge of particular
galaxies is rather better, and we emphasize that we are studying
generic properties of thin disc dynamos.

Taking typical galactic values, we can estimate $R_\alpha=O(1)$,
$R_\omega=O(10)$ (i.e. $D = O(10)$). We adopt values $R=10$~kpc,
$r_h=1, h_0=350~{\rm pc}$. This gives $h \approx 500$~pc at
$r=R=10$~kpc. This gives a time unit of $0.78$ Gyr.
We use a fixed integration timestep of approximately $0.04$ Myr.

We superimpose on this rather standard dynamo model the injection of
random magnetic fields of rms strength $B_{\rm inj0}$ with energy density
comparable to that
of the turbulent motions, at
discrete time intervals and at a number of randomly determined
discrete locations. We take the interval between injections to be about
$10^7$~yr -- see \cite{metal12} for details. (In the context
of our models this time can be interpreted as a convenient interval
at which to maintain/renew the injection of small-scale field.
This interval is close to the turnover time of vortices
in the interstellar turbulence, which in turn determines the
time scale of small-scale magnetic field evolution.) The new feature of this
paper is that injection only occurs within a spiral
pattern representing the material spiral arms. This pattern is
assumed to rotate rigidly with the pattern speed $\omega_{\rm P}$.
We choose $\omega_{\rm P}$ such that the corotation radius $r_{\rm corr}=0.7R$
in each case.
For simplicity, the injection rate falls discontinuously to zero at
the boundaries of the arms. The shape of the arms, and of the
centrally enhanced central injection region, can be seen in, e.g.,
Fig.~\ref{evol}. This is combined with an enhancement of the
injection field strength near the galactic centre (an increase by a
factor $2$ at the centre, decreasing to unity at $r=0.1$ or $1$~kpc
in dimensional units). This is a  token
recognition of the
increased star formation expected to occur in the central
part. We stress that $B_{\rm inj0}$ should be considered only as a
proxy for the typical field strength generated by the small-scale
dynamo action associated with star-forming regions. Thus we allow it
to take a range of values, to compensate for inherent deficiencies
in the model.

Our expectation is that as the material arms sweep through the
ambient gas (or v.v. depending on the position relative to the
corotation radius), outside of the regions where field is injected
the differential rotation will be able to organise the small-scale
field into large-scale. Inside the arms, both small- and large-scale
field will be present. Effectively, the ISM will be "reseeded" by
the ongoing injections.

The dynamo equations are integrated on a $537\times 537$ Cartesian
grid which is just large enough to provide a ``dead zone'' around
the dynamo active region -- see e.g. Fig.~\ref{evol}.
To enable reproducibility and for ease of inter-comparison, 
the same sequence of pseudo-random numbers was used in each simulation.

\section{Arm-interarm contrast of dynamo governing parameters}

In this Section we use the parametrization of the arm-interarm contrast for the
dynamo governing parameters introduced by Shukurov \& Sokoloff
(1998). As the $\alpha$-coefficient is larger in the interarm
regions, whereas the kinetic energy density of turbulence is larger
in the arms, the turbulent magnetic diffusivity can only be weakly
affected by the spiral pattern. In turn, these parameters are
reduced via more or less conventional scaling to the turbulent rms
velocity $v$, the basic scale of galactic turbulence $l$, the scale
height of the ionized gaseous galactic disc $h$, the galactic
angular velocity $\Omega$ and the gas density $\rho$. The
arm-interarm contrast is estimated as $v_{\rm a}/v_{\rm i} \approx
2$ for $v$, and $\rho_{\rm a}/\rho_{\rm i} \approx 3$ for the gas
density (indexes $a$ and $i$ stands for arm and interarm region,
respectively), based on available observational data for the Milky
Way (Rohlfs \& Kreitschman, 1987) and M~51 (Garcia-Burillo et al.,
1993, see also Fletcher et al. 2011). The contrast in turbulent energies
is estimated as $E_{\rm a}/E_{\rm i} \approx 10$.
The scale heights $h_a$ and $h_i$ were
taken to be equal, because the sound crossing and passage times
for the spiral density wave were estimated to be almost the same, and so the
density wave affects the mean hydrostatic equilibrium
only.
The estimate of the
turbulent diffusivity contrast is based on the assumption that the
correlation time scales as the inverse supernova rate, which gives
$\eta_{\rm a}/\eta_{\rm i} \approx 1$. The correlation length $l$ is
identified with the radius of supernova remnants. Shukurov \&
Sokoloff (1998) take  a conventional estimate $l_{\rm a} / l_{\rm i} \approx 0.5$. Assuming
that $\Omega$ is the same in arms and in the interarm regions, they
obtain the estimate $\alpha_{\rm a} /\alpha_{\rm i} \approx 0.25$.
We accept these estimates as representing the current
state of knowledge.

\begin{figure*}
\centering
\includegraphics[height=0.90\textwidth]{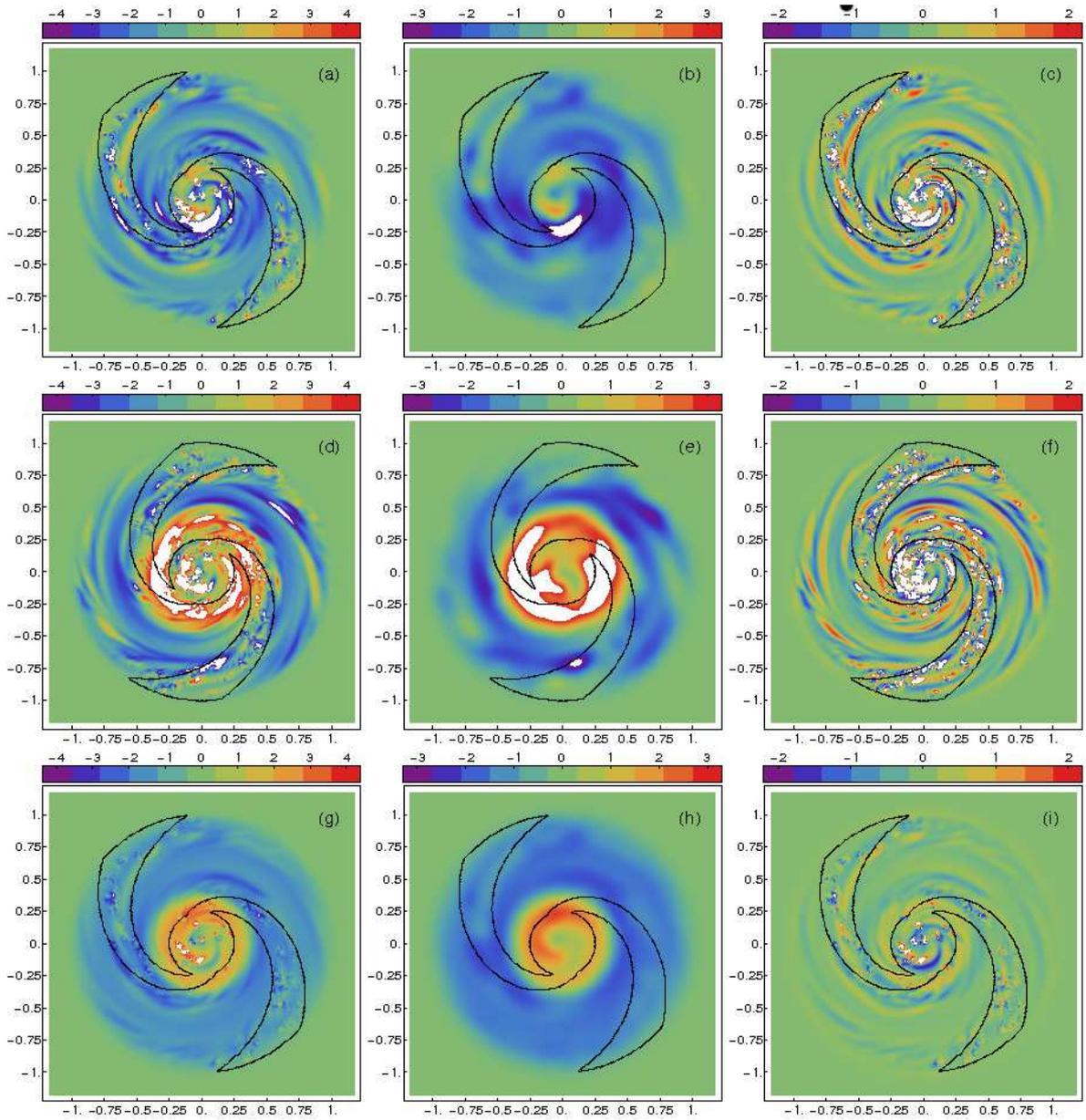}
\caption{Colour coded images of our model galaxies: from top to the bottom
models 75, 76 and 77, from left to right: azimuthal component of
magnetic field, azimuthal component of large-scale magnetic field
and azimuthal component of small-scale magnetic field at
approximate times $13.2$~Gyr. The large-scale field
is averaged over $1$ kpc. The continuous contours delineate the regions
("arms") where field is injected;  these rotate rigidly,
with pattern speed such that the corotation radius is at
fractional radius $r\approx 0.7$.}
\label{colour_evol}
\end{figure*}

As context for our main calculations, we performed dynamo
simulations with $\alpha$ increased by a maximum factor of about 3.3
between the arms with  a smooth quasi-parabolic profile of $\alpha$.
With $R_\alpha = 1$, $R_\omega =4.5$, the dynamo is only slightly
supercritical. Here we were guided by the experience of Shukurov (1998)
that weak dynamo action is more favourable for the production of interarm fields by this mechanism.
For a weak initial field it is possible to see a
slight enhancement of the (regular) field within the interam region
(we do not display this result in detail here).
This run
started with a random field with rms strength $10^{-6}$ (but
no further field injections).
The resulting magnetic configuration
becomes more complicated for a more substantial initial field (with
a strength of order of equipartition), but again there are no very
pronounced features localized in the interarm regions.
Apart from
these small azimuthal variations, there is also a radial field
reversal, but nothing like the magnetic arms of NGC~6946. Note that
for a weak initial field the time to saturation is about 17~Gyr (see
also the discussion in Moss \& Sokoloff (2012a)). This time becomes
much shorter and realistic with a strong initial field (cf. Arshakian et al.
2009). This
mechanism, taken alone, does not seem particularly promising.
We note that a more elaborate model of this general type by Chamandy et al. (2013a)
produces potentially more interesting results.

We conclude that special variations of dynamo governing parameters,
e.g. as considered by Moss (1996), can in principle give magnetic
arms, but requires some fine tuning of parameters (see also Rohde et
al. 1999). We deduce that variations which correspond to the naive
estimates of Shukurov \& Sokoloff (1998) fail to produce pronounced
magnetic arms. It follows that to explain the magnetic arms in the
framework of the model under consideration, we need something more
than just modulations of $\alpha$ by spiral arms, such as continuous
small-scale field injections, modulated by spiral structure.
We emphasize that azimuthal variations of dynamo quantities (here $\alpha$)
are considered only in this Section, for illustrative purposes.

\begin{figure*}
\centering
\includegraphics[height=0.6\textwidth]{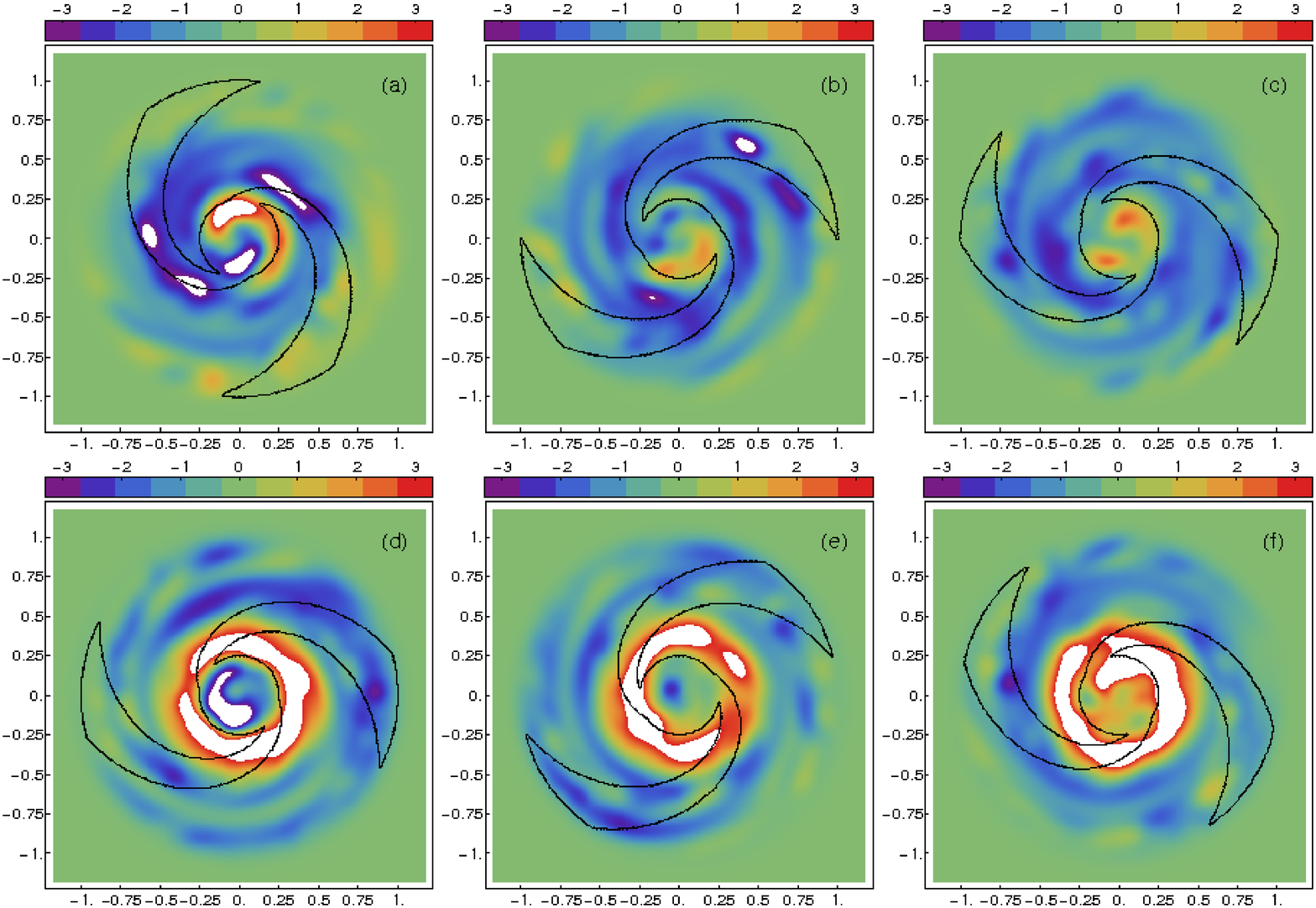}
\caption{Colour coded figures for the azimuthal component of the
mean magnetic field at time evolution from left to right: $11.7$, $12.5$ and $12.9$ ~Gyr. Fields are averaged
over lengths $1$ kpc. The top row shows model 75, the lower model 76. The continuous contours delineate the regions
("arms") where field is injected;  these rotate rigidly,
with pattern speed such that the corotation radius is at
fractional radius $r\approx 0.7$.} 
\label{mean_evol}
\end{figure*}

\section{Magnetic configurations with field injections modulated by spiral arms}

Moss et al. (2012) included  the injection of small-scale magnetic fields as a proxy for the effects of the dynamical
star-formation
process in their model for the evolution of magnetic fields in disc
galaxies. They
assumed that the star-formation rate governs the rate of supernova
explosions, which in turn drive the turbulence of the gas in these
galaxies and are the main source of small-scale magnetic fields via
small-scale dynamo action. In this paper, we simulate the
evolution of the large-scale magnetic field,
by assuming the injection rate of small-scale field,  of
strength of order the equipartition strength,
to be high only in the material spiral arms
-- since more gas resides in the arm regions (and hence the
star-formation rate is higher there) compared to the interarm
regions.
In the following discussions, we do not allow any arm/interarm variation of the unquenched alpha-coefficient,
in order to isolate clearly the effects of field injection in the arms.

We discuss some representative models; the adopted values
of $R_\omega$ give a rotational velocity of around $200$ km s$^{-1}$,
comparable to that in the Milky Way.
In the first column of Fig.~\ref{evol}  we show our reference model (model 75) for which
$R_\alpha = 3$, $R_\omega = 12$, $h_0=350~{\rm pc}$  (so $h\sim 500$ pc at
radius $10$ kpc) and $B_{\rm
inj0} = 10$, from time $11.7$~Gyr to $13.2$~Gyr, corresponding to the present day.
(In this, and other models discussed, the corotation radius is about
$7$~kpc, i.e. $r=0.7$.) It is clearly visible that the magnetic
field is more regular between the spiral arms than in them.
We also show in Fig.~\ref{early} snapshots of the earlier
evolution of model 75. These two Figures demonstrate a gradual 
development of large-scale magnetic field in the interarm regions. 
Indeed, we can see in these two Figures
the progression in model 75 
from a largely disorganized field (localized in the arms),
to a fully developed global field with large-scale structure.
By playing with the dynamo governing parameters we can modify the
result. A reasonable increase of differential rotation (to
$R_\omega= 20$) adds a global magnetic reversal in the central part
of the galaxy (model 76, Fig.~\ref{evol}, second column) while a lower injection
rate gives some regular field in the material arms (model 77,
Fig.~\ref{evol}, third column).

\begin{figure*}
\centering
\includegraphics[height=0.30\textwidth]{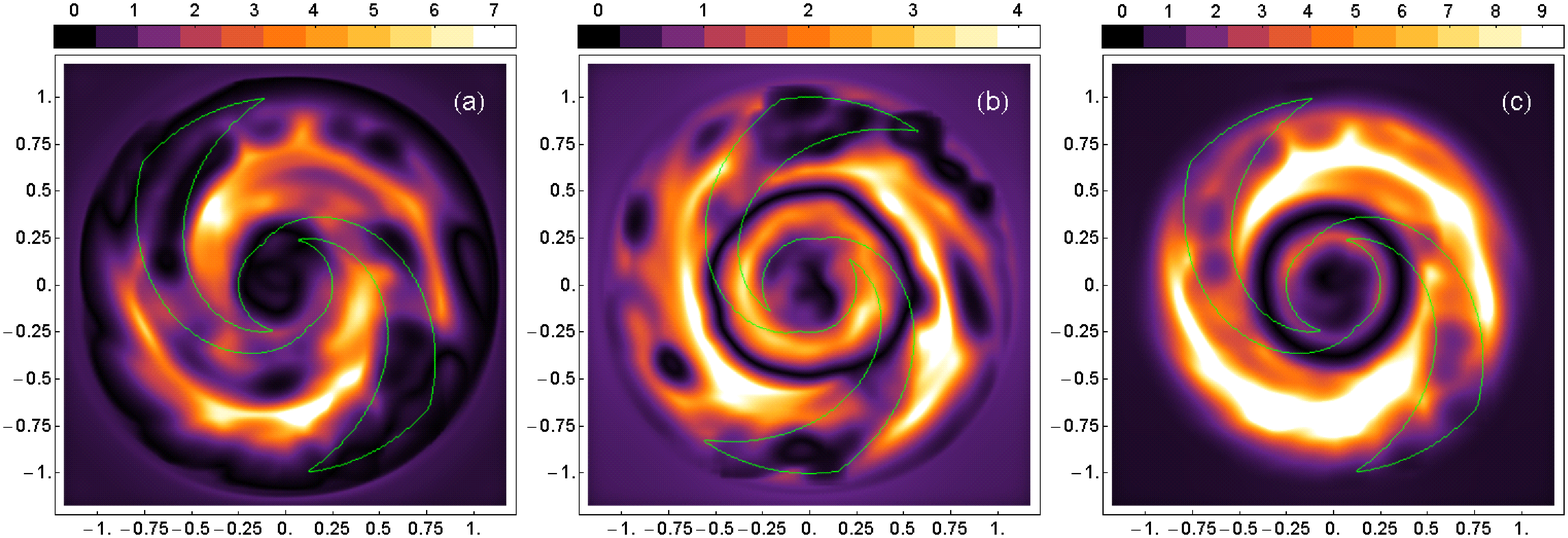}
\caption{Colour coded images showing the ratio of mean (large-scale) to r.m.s. for the galaxies: from left to right
models 75, 76 and 77 at
approximate times $13.2$~Gyr. The fields are averaged over $1$ kpc. The continuous contours delineate the regions
("arms") where field is injected;  these rotate rigidly,
with pattern speed such that the corotation radius is at
fractional radius $r\approx 0.7$.}
\label{meantorms}
\end{figure*}

Decomposition of magnetic field into large-scale and 
small-scale components can be performed in various ways
and it is far from clear in advance
which particular way is more physically meaningful (cf. Gent et al. 2013).
As an example, we present in Fig.~\ref{colour_evol} the total, large-scale and small-scale azimuthal magnetic field obtained by  applying a Gaussian filter with 
width $\sigma=500$ pc to the
computed magnetic field. We can compare panels of the second column of
Fig.~\ref{colour_evol}, showing the large-scale field, with the panels
of the third column which show the small-scale field which
in our model coincides with the position of the material arms.
In addition Fig.~\ref{mean_evol} shows also some earlier stages of the
evolution of azimuthal component of the mean field in models 75 and 76. 
Models 75 and 76 exhibit 
dynamical evolution of magnetic arms which can either be located between the 
material arms or cross them, 
whereas it can be  seen clearly that model 77 has  an almost completely 
axisymmetric large-scale magnetic field.
There is no fixed certain correlation between the large-scale  
structure of the magnetic  fields and the material arms. 
The
arm/interarm contrast is clearly shown  by the ratio of large-scale field
 to the root-mean-square of the  fluctuations (see Fig.~\ref{meantorms}).

Note that the reference model 75 shows many local reversals, which
can be compared with local reversals detected in the Milky Way (Van
Eck et al. 2011). In contrast, model 76 shows a global
reversal. A plausible argument for larger differential rotation
($R_\omega$) favouring reversals can be made as follows. To get
reversals we need relatively strong magnetic fields to be present at
the early stages of galactic evolution. This is (as we have seen)
more likely for stronger initial fields with ``enough'' lumpiness.
With too much lumpiness, the initial conditions are near-uniform,
which may, for suitable choices, give no reversals. Larger
$R_\omega$ means that the initial inhomogeneities grow faster,
without diffusing radially -- they get stretched into rings. There
is more chance of the subsequent radial merger of these rings giving
extended radial regions with opposing $B_\phi$ - i.e. reversals.
Further discussion is given in Moss \& Sokoloff (2012b).

 We see a marginal effect (visible clearly only away from corotation,
in certain models), that the magnetic field upstream of the arms
($r_{\rm corr}=0.7R$) is slightly more disordered than that downstream.
Note that the downstream side of the arm changes between the regions inside
and outside of corotation.

\begin{figure}
(a)\includegraphics[height=0.33\textwidth]{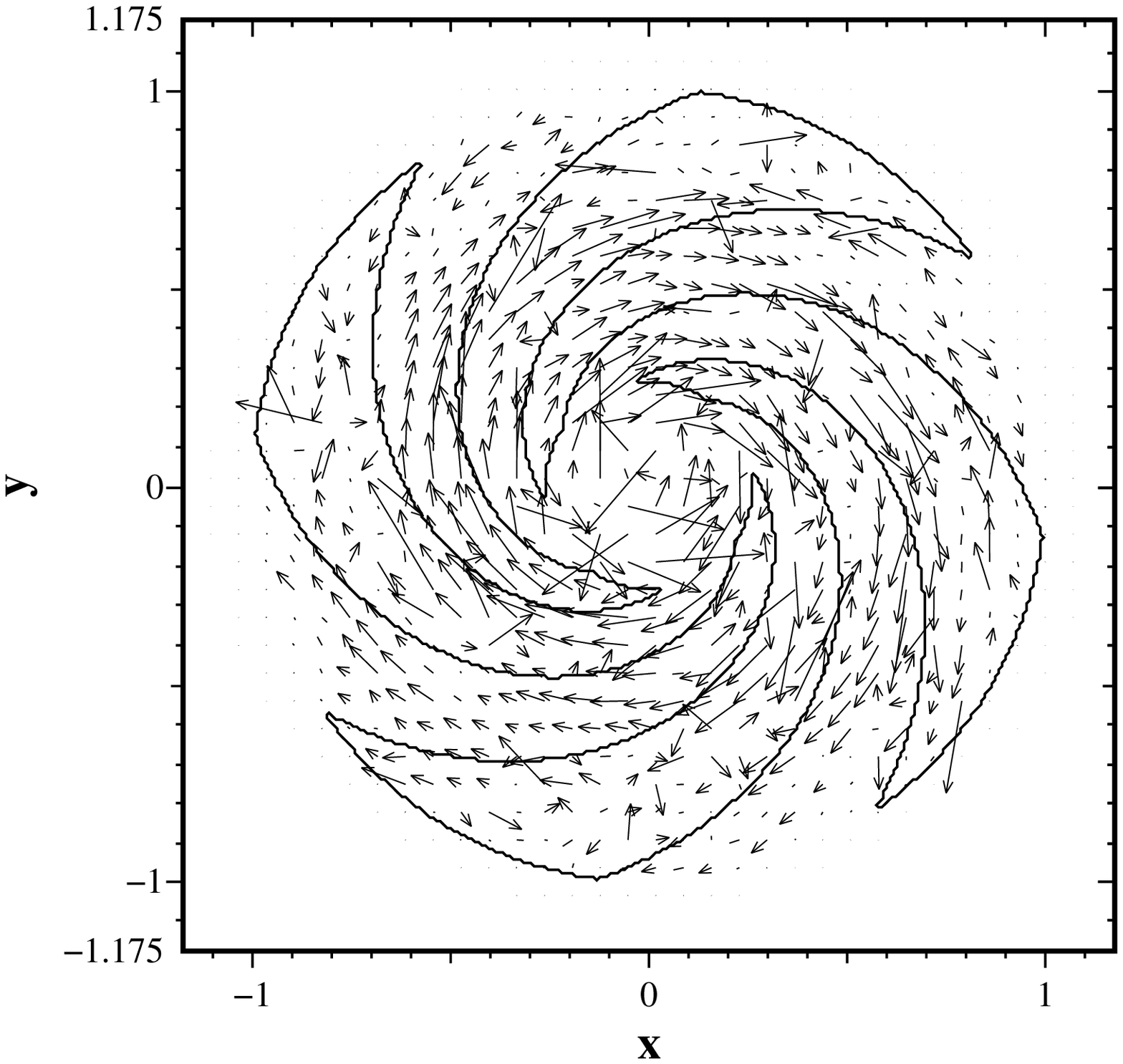}\\
(b)\includegraphics[height=0.33\textwidth]{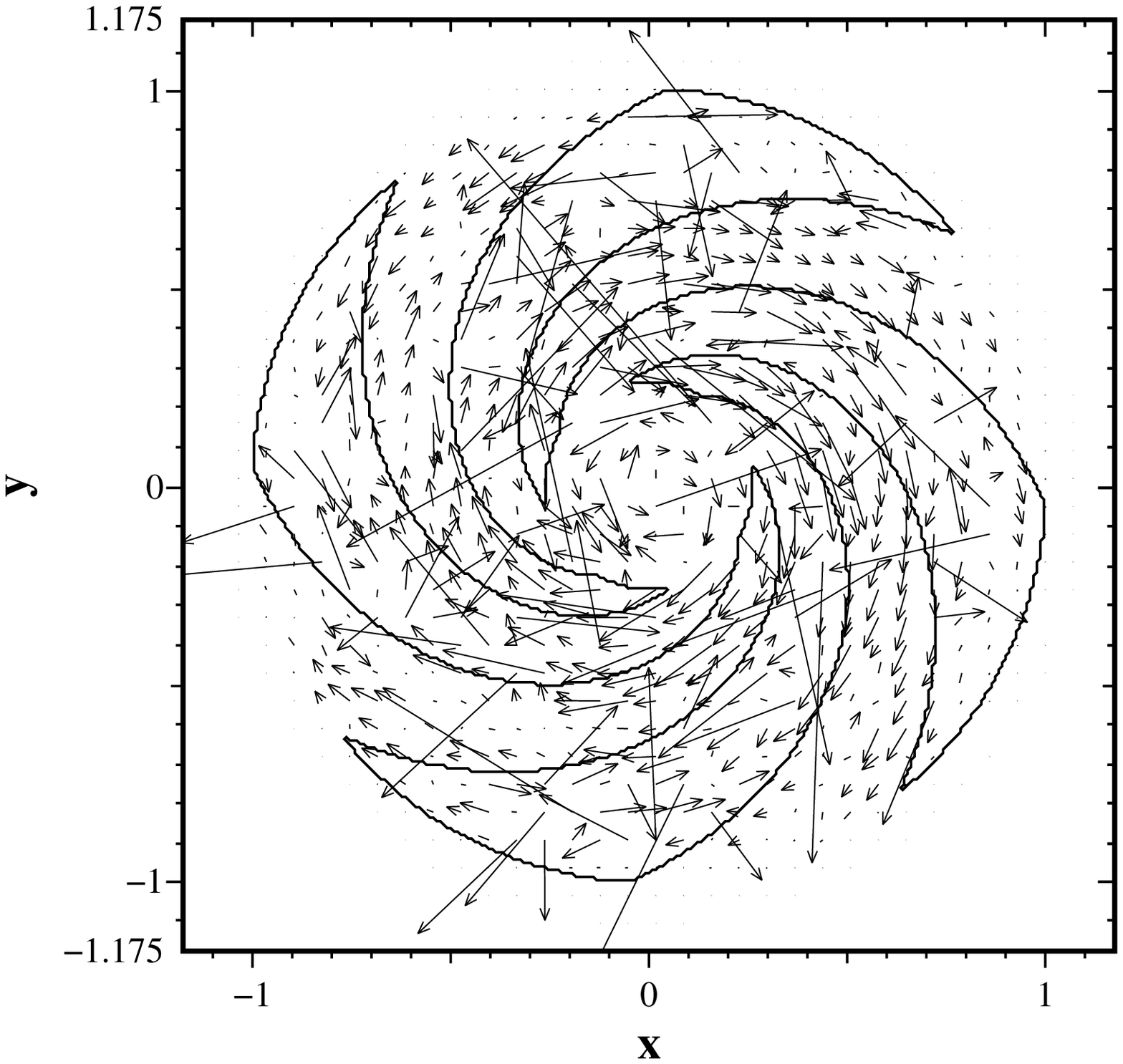}\\
(c)\includegraphics[height=0.33\textwidth]{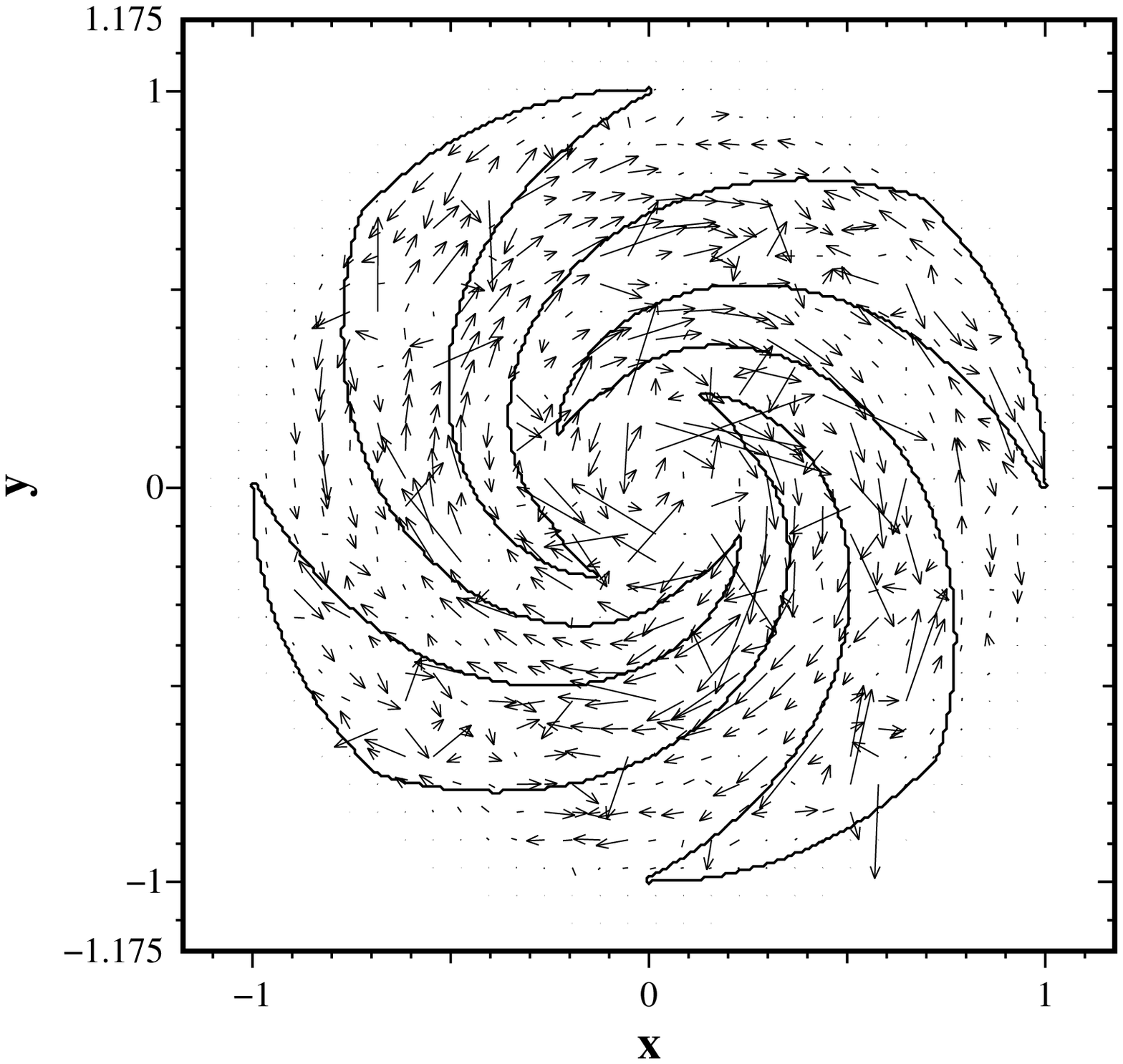}\\
(d)\includegraphics[height=0.33\textwidth]{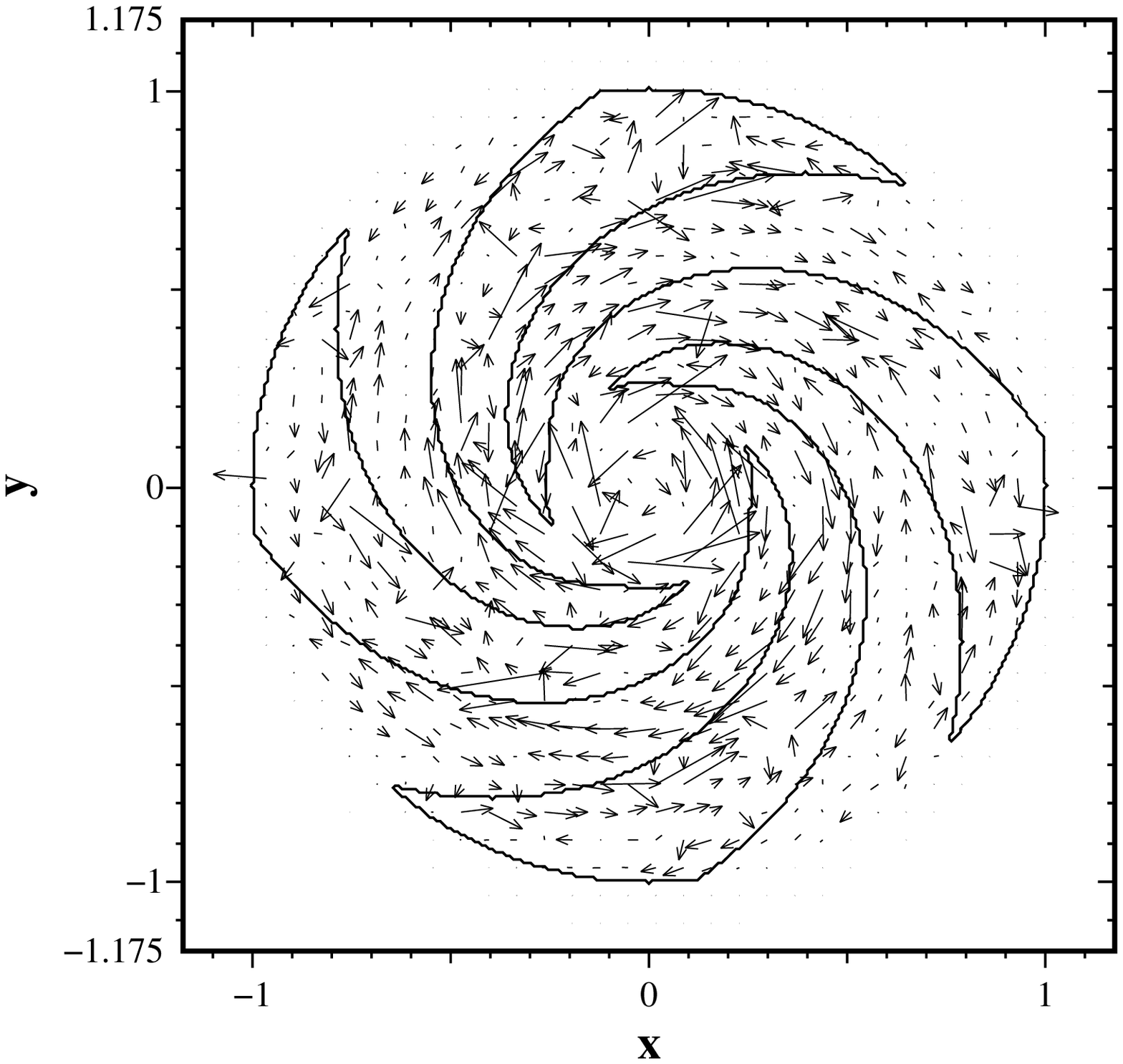}
\caption{Magnetic arms in a 4-armed galaxy (model 402) at
approximate times $11.7, 12.5, 12.9 ~{\rm and}\ 13.2$~Gyr. The dynamo
parameters are the same as for model 75. The contours delineate the material arms.} \label{develop}
\end{figure}

We also investigated a similar model with 4-armed spiral structure,
and show, in  Fig.~\ref{develop}, snapshots of the magnetic field
at later times for a model
with the same dynamo parameters as model 75.
The magnetic arms situated
between the material arms now look much less pronounced than in the
corresponding 2-armed model (model 75), plausibly because there is
now not enough time for an ordered field to arise in the interarm
regions, before the next arm comes along and disturbs it. Again,
when we vary the intensity of the field injections, we find that the
magnetic arms are smoothed out if less small-scale field is
injected.

The evolution of magnetic patterns in model 75 is presented in
Fig.~\ref{evol}. We see that details of the magnetic pattern shape
are variable on timescales comparable with the galactic rotation
period. A magnetic arm (quite smooth in Fig.~\ref{evol}a) is much
sharper in Fig.~\ref{evol}b. The main point is that the random
injection of the small-scale magnetic field, which mimics the role
of supernovae in star-forming regions in generating small-scale field,
is important and determines the instantaneous shape of the
magnetic configuration. This is an intrinsic property of our dynamo
model. A further, more artificial, effect is that our injections of
field occur simultaneously at discrete intervals in time. The
anomalously large isolated B-vectors visible in Fig.~\ref{evol}d,e
probably arise from such an immediately previous injection, coupled
with the intrinsically random distribution of the injection sites,
strengths and the choice of the locations at which vectors are plotted. The
contemporary state of observational studies of galactic magnetism does
not support or reject this conclusion, because we have snapshots of
magnetic patterns in a very few galaxies only. However we stress
that this conclusion could be supported by observations of a
representative sample of spiral galaxies. The prediction is that we
would expect to see a rich variety of magnetic patterns in such a
sample. Fine tuning of parameters of a particular dynamo model to
mimic, say, the magnetic patterns in NGC~6946 is less useful in
the framework of our dynamo model, because snapshots of magnetic
configurations over relatively short time intervals (on galactic
time scales, of course) can differ quite significantly.

Some dynamical simulations (e.g. Wada et al. 2011) of the evolution
of spiral galaxies show that material arms dissolve and reform on
relatively short timescales, typically several rotation periods. In
contrast, in the models discussed above, the arms are permanent
features. To assess the importance of this simplification, we ran a
model with the parameters of model 75, in which the position of the
arms changed discontinuously and at random at intervals of about $5
\times 10^8$~yr. In this case, at least, the field at large times
was not generically different to that shown in Fig.~\ref{evol}. Of
course, immediately after such a jump there will be some
differences, such as temporary non-coincidence of disordered field
and material arms, but this is short-lived. If the jump is by an
angle near $0^\circ$ or $180^\circ$, there is little gross effect, even
immediately.

Additionally, we ran simulations with parameters as model 75 
with the equipartition
field strength in the material arms increased by ca. 25\% and ca. 70\%. This
(maybe unsurprisingly) resulted in a modest increase in magnetic energy,
but caused very little difference to the overall field morphology.
In particular, the marked contrast between arm and interarm regions remained.

\section{Discussion and conclusions}

We have presented a galactic dynamo model that attempts to explain 
the phenomenon where the degree of uniformity of the magnetic field is higher 
in the interarm regions.
This work is largely motivated by observations such as those
of IC~342 (Krause 1993) and the prototypical case of NGC~6946 (Beck
2007) that magnetic arms can be situated between material arms.
M~51 is one of the rare cases where magnetic and material
spiral arms almost coincide because
of compression (Fletcher et al. 2011). Density-wave shock fronts
cause velocity perturbations which are not included in our model.
The key feature of our model is that it includes small-scale
magnetic field injections from small-scale dynamo action, associated
with strong turbulence in star-forming regions that are
predominantly found in spiral arms and near the galactic centre.
 When the gas in which this small-scale field is embedded leaves the
material arms, conventional large-scale dynamo action can use it as an
ongoing ``seed'' to produce ordered (regular) field between the material 
arms.
We have demonstrated that modulation of the small-scale magnetic
field injection rate allows a mimicking of the phenomenology of
ordered interarm fields. 
We obtain snapshots of magnetic fields that look broadly similar to the
observational plots of the polarized intensity of many galaxies including
NGC~6946 (Beck 2007) -- contemporary ordered fields in model 75 are
situated between the material arms. But we note that in model 77 the
ordered field 
almost 
fills the whole interarm and arm space, and does not
form distinct magnetic arms.

\begin{figure*}
\centering
\includegraphics[height=0.43\textwidth]{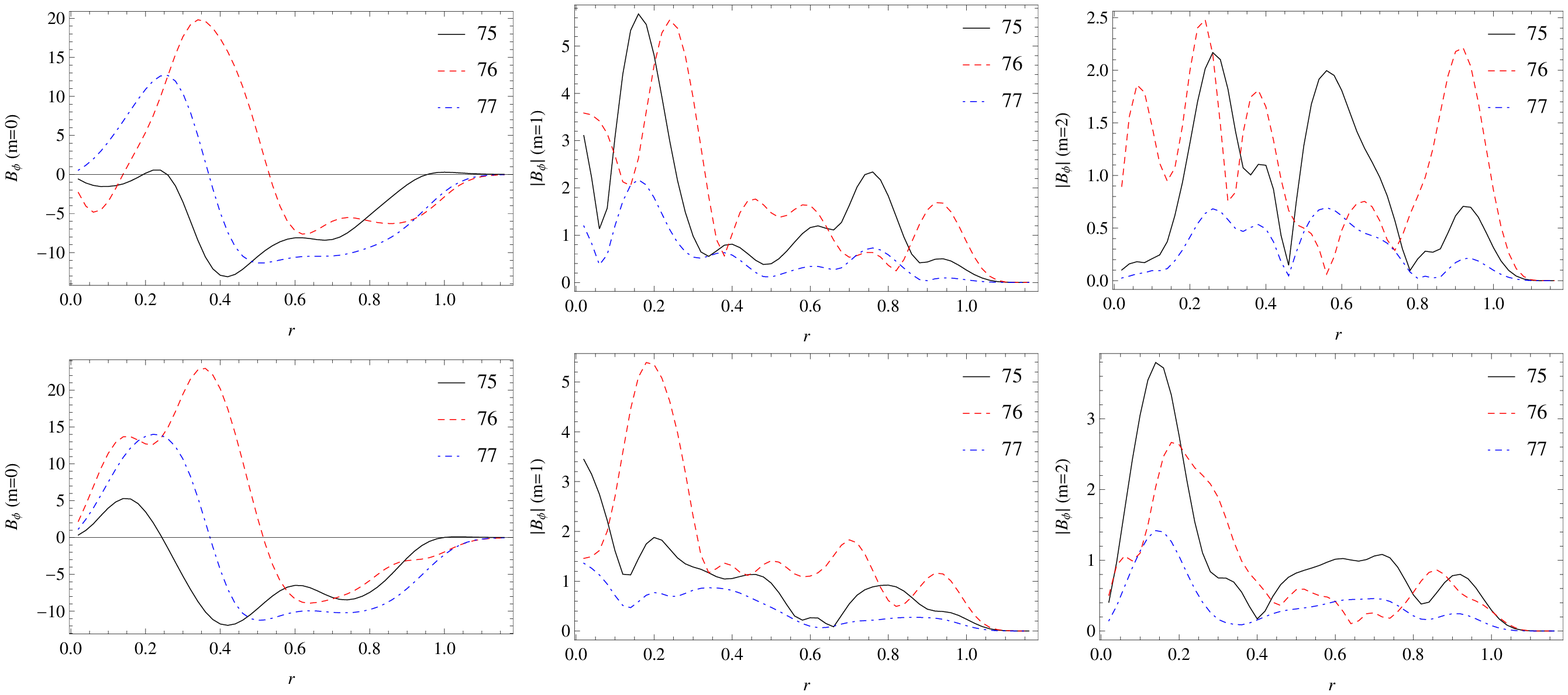}
\caption{The amplitudes of azimuthal modes of large-scale of $B_\phi$ vs radius: $m=0$ - left hand column; $m=1$ - middle column; $m=2$ - right hand column. The
first row is at approximate time $12.5$~Gyr and the second row at $12.9$~Gyr. }
\label{modes}
\end{figure*}

A more realistic approach might, e.g., have the field injection
decrease more slowly and smoothly away from the arms. This could be expected to
result in narrower regions of regular field (``magnetic arms'').

Our results are generally consistent with a process in which disordered
field injected within the material arms continually finds itself
in the interarm regions, as the material arms move relative to
the gas in the disc. Once outside of an arm, large-scale order is imposed by
the action of the global dynamo (which is now unperturbed by field injections)
together with the stretching associated with the differential rotation.
Of course, the large-scale dynamo action occurs within the material arms also,
but there its effects are obscured by the injected random field.
We believe that this is the key mechanism of our model.

We note that the magnetic structures are more tightly wound than the material arms, and thus intersect them.
We did not see a strong influence of the corotation radius in our models,
except perhaps that the magnetic arms may cross the material arms near or
just inside the corotation radius.
This is visible in models 75 and 76, but not in model 77 (Fig.~\ref{evol}),
where discrete magnetic structures do not occur.
As the magnetic structures are quite broad
in our models, it is difficult to be more precise. This also means that the
large-scale field seems to be more widely distributed in radius than in the
models of Chamandy et al. (2012).

A quantification of the phenomenon in terms of the concepts of
mean-field theory is faced with the problem of how to
distinguish mean and large-scale features.
We have made a preliminary attempt to contrast mean field and fluctuations in Figs.~\ref{meantorms}.
The breadth of the
material arms is quite small in comparison with the length of the
ring centred at the galactic centre at a typical radius; the
global perturbation to the basic axisymmetric structure is
small. A consequence is that the magnetic arms give only a weak
$m=2$ (or $m=1$) mode with amplitude changing relatively quickly with time,
when decomposed into azimuthal modes (see Fig.~\ref{modes}).
It seems more satisfactory to say
that the magnetic arms should be considered as mesoscale phenomenon
of the magnetic field configuration. In any case, the magnetic arm
phenomenon cannot be considered as an excitation of one of just one or two
of the lower non-axisymmetric modes of the mean field. Rather it is a coherent
structure which can be represented by many coherent Fourier modes.

In the context of the arm-interarm contrast, 
Fig.~8 shows that the magnitude of the large-scale field in the arms is not very
different to that in the interarm regions in all the models, 
in contradiction to observations --
our models predict a significant
large-scale field in the arms, almost as strong as in the interarm
regions. There are several
mechanisms that can reduce the expected polarized emission in the arms:
tangling by SNR shocks and shear motions, Faraday depolarization and
averaging effects due to the limited beam size.
Another possible resolution of this point is that we were unable to include
consistently an enhancement of turbulent diffusivity in the arms, as we were
unable to represent gradients satisfactorily in the code. However,
experiments with enhanced diffusivity, without inclusion of gradient
terms, suggest that this mechanism is capable of alleviating the problem
by reducing the large-scale field in the arms.

Note that our models neglect Faraday depolarization effects and hence should be
compared only to radio polarization maps at high frequencies where
Faraday depolarization is small, typically at wavelengths of $\lta 6$ cm where
most polarization observations were performed (as for for NGC 6946, see
Beck 2007). At longer wavelengths, polarized radio emission traces not
only large-scale magnetic fields, but also the amount of Faraday
depolarization.

We have emphasized models with relatively well-developed
 magnetic arms 
-- our main purpose in this paper is to demonstrate "proof of
concept", rather than to present an exhaustive survey of parameter space.
However we note that both a weaker strength of injected field,
and stronger differential rotation ($R_\omega$) make magnetic arms less pronounced 
in the sense that magnetic field in the arms becomes less disordered, and the large-scale
field becomes more homogeneous globally -- see Fig.~\ref{evol}.

\begin{figure}
\centering
\includegraphics[height=0.75\textwidth]{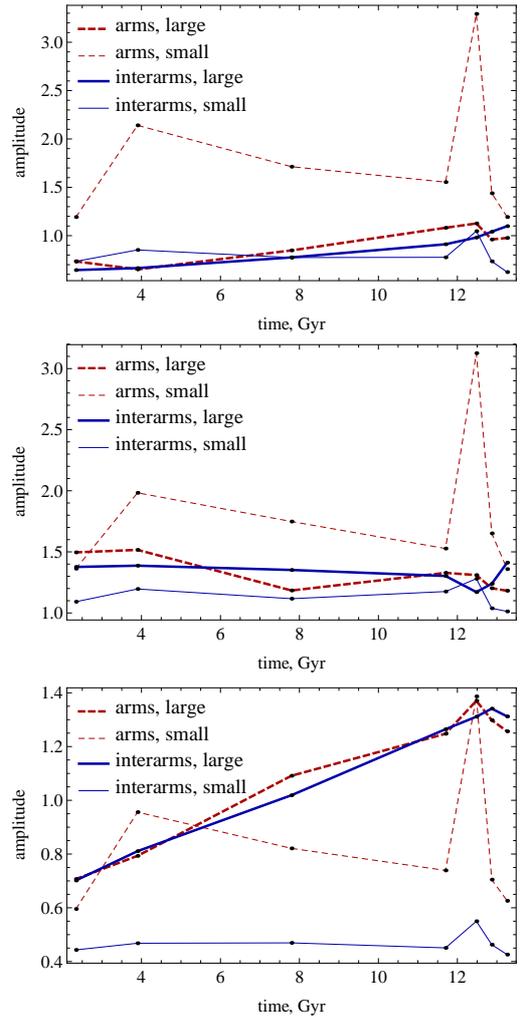}
\caption{Quantification of arm/interarm contrast for large-scale (thick lines) and small-scale (thin lines) fields for material arms (red dashed
lines) and interarm regions (blue solid lines), for models  75, 76, 77 (from top to bottom). The peak in small-scale field at approximate time
12 Gyr as well as other details of small-scale magnetic field distribution are probably caused by the chance coincidence of an injection event with the 
analysed snapshot of the field.}
\label{Gaussfilter}
\end{figure}

We calculated the averaged amplitude of the large-scale magnetic field 
and rms of the small-scales over arm and interarm regions separately.
The panels of Fig.~\ref{Gaussfilter} should be
compared with Fig.~\ref{evol}, where the arm-interarm
contrast looks pronounced for the models 75 and 76,
but very much weaker for model 77. We see clearly from
Fig.~\ref{Gaussfilter} that the small-scale field in the arms is
noticeably larger than the large-scale field in
the models 75 and 76, and is substantially weaker
in the later stage of the evolution
of model 77. As a result, the magnetic field in the arm region
looks disordered for models 75 and 76, and
quite ordered for model 77. We verified that this quantification
remains stable for filter widths up to
$\sigma=1$ kpc, while larger filter widths become more comparable with
the arm width and the contrast is smoothed out.

From the above analysis our result looks quite
straightforward: the magnetic field becomes less
ordered in the regions where the field injections are stronger.
The outcome is nevertheless nontrivial, because we have
seen that enhanced dynamo action alone does not lead to a similar contrast.

We stress however that the analysis used to generate Fig.~\ref{Gaussfilter}
does not reproduce the visual impression
of Fig.~\ref{evol} in that the large-scale field in the arms is not
significantly weaker than in the interarm regions. The reason for this behaviour
seems to be that the large-scale dynamo operates throughout the disc,
and produces a global scale field. In the arms, small-scale field is
added, which is converted to large-scale field in the interarm
region before entering the next arm. It is not clear in advance how this is
related to the processing of the observational data. In
other words, interpretation of the results depends on the algorithmic
distinction between the concepts of large-scale,
mean, regular, ordered and other characteristics of the
 visual impression of the magnetic field, which are applied in
interpretation of observational data.
Progress in this direction
is obviously strongly required, but it is however
obviously far beyond the scope of this paper.

Studies in the far-UV, optical and H$\alpha$ integrated light
in regions of recent star formation of NGC~5236 (M~83)
(Lundgren et al. 2008; Silva-Villa \& Larsen, 2012) allow estimates
of the  star formation rate to be made.
It was found that the surface star formation rate
densities are higher in arms than that in the interarm regions by
approximately $0.6$ dex, and that the star formation rate changes along
the spiral arm, being higher in the leading part than in the trailing
part of the arm (Martinez-Garcia et al. 2009; Silva-Villa \& Larsen
2012). As was shown in our simulations
(models 75j, 76k in Fig.~\ref{evol}), the higher injection rate of
turbulent magnetic field suggested to be associated with a
higher star formation rate
in the arm compared to the rate in the interarm regions, leads to a
less ordered field in the arms and large-scale regular fields in the
interarm regions. The
offset of the star-formation rate in the arm and interarm regions is
thus able to explain the contrast of the regular magnetic field in those
regions.
The variation of the star-formation rate along
the spiral arm may have a specific imprint on the ordering of the
magnetic field along the arm. High resolution and high sensitivity
observations of a dozen or so nearby galaxies with the SKA are needed
to address this issue. We intend to study the effect of the variation of
the star formation rate on regularity of the field along a spiral
arm in subsequent papers.
We note that some of our models possess large-scale field reversals of
the type believed to be present in the Milky Way (e.g. van Eck et al.
2011; Farrar \& Jansson 2012), while they seem to be rare or absent in external galaxies.
Polarisation observations with forthcoming radio telescopes like the
Square Kilometre Array (SKA) and its precursors can help elucidate this
issue (Beck 2011).

Following the temporal evolution of the magnetic arms, we find that they are
quite variable entities and that their shape can change
significantly from one snapshot to the other. An arm which looks
quite diffuse in one snapshot may become concentrated in another,
and the relative position of magnetic and material arms can vary in
time -- see e.g. Fig.~\ref{mean_evol}.
In this sense magnetic arms are  - similarly to material
spiral arms - much less stable than large-scale magnetic reversals,
which are sometimes present, e.g. in model 76.

Our models have the potential to describe a wide range of types of galactic 
magnetic field structures. We can note that
situations such as shown in model 77 seem to be rarely 
observed.
The only possible candidate is M31 where the large-scale field seems
to fill both arm and interarm regions (Berkhuijsen et al. 2003), but
the  high inclination of M31 does not allow a definite statement.

An increase in the number of material arms tends to produce
less axisymmetric global magnetic structure (Fig.~\ref{develop}) 
and, as expected,
decreasing the injection rate also makes magnetic arm structure weaker.
We note that some studies suggest that the Milky Way has a thicker disc than
adopted in our models (Schnitzeler 2012), and that the disc may not be flared
(Lazio \& Cordes 1998).
We stress that we are not trying specifically to model the
Milky Way. However we can note that a thicker disc would mean the dynamo was
more efficient, and so similar results to ours could be obtained for
smaller values of the dynamo numbers -- while staying within
the intrinsic uncertainties in our knowledge of these quantities.
This would not affect significantly the
operation of our proposed mechanism for magnetic arms.
(Of course, in a model that includes an explicit
$z$-dependence, rather than our two dimensional, vertically averaged,
approximation, the thickness of the disc is a significant input quantity.)
In general, we believe our results will be quite robust with respect to
plausible changes in galactic parameters, such as disc thickness
and flaring, within the uncertainty of the basic dynamo parameters.

In conclusion, we have demonstrated that the process studied,
 namely the contrast in star formation rate between arm and interarm
regions, can make a
major contribution to the large-scale ordering of field in the 
arm/interarm regions, including the phenomenon of magnetic arm generation 
and the presence of magnetic arms situated between material arms, 
for suitable choices of parameters.
We have examined in isolation the simplest modelling of
field injection,  but a more realistic modelling would certainly include
other mechanisms, with corresponding
modulation of our preliminary results.

\begin{acknowledgements}
The authors thank the referee, Anvar Shukurov,
for making a number of comments that led to substantial improvements to the paper.
DM, DS and RS are grateful to MPIfR for hospitality. The paper was
partially supported by RFBR under grant 12-02-00170. RB acknowledges
support by DFG grant FOR1254. The work was also partially supported by
DFG project number Os 177/2-1.
\end{acknowledgements}

\end{document}